\documentclass[preprint,11pt]{article}

\usepackage{color}
\usepackage{graphicx}
\usepackage{amssymb}
\usepackage{amsmath}
\usepackage{natbib}
\usepackage{algorithm}
\usepackage{algorithmic}

\newcommand{\R}{\mathbb{R}}

\newcommand{\bC}{\mathbf{C}}

\newcommand{\bT}{\mathbf{T}}
\newcommand{\bX}{\mathbf{X}}

\newcommand{\bZ}{\mathbf{Z}}

\title{Modeling and simulating depositional sequences using latent Gaussian random fields}
\author{Denis Allard\\ 
INRAE, BioSP, Avignon, France\\
Paolo Fabbri\\
University of Padua, Padua, Italy\\
Carlo Gaetan\\
Ca' Foscari University of Venice, Venice, Italy
}

\begin{document}
\medskip
\date{\today}
\maketitle

\begin{abstract}
%% Text of abstract
Simulating a depositional (or stratigraphic) sequence conditionally on  borehole data is a long-standing problem in hydrogeology and in petroleum geostatistics.  This paper presents a new rule-based approach for simulating depositional sequences of surfaces conditionally on  lithofacies  thickness data. The thickness of each layer is modeled by a transformed latent Gaussian random field allowing for null thickness thanks to a truncation process. Layers are sequentially stacked above each other following the regional stratigraphic sequence.  By choosing adequately the variograms of these random fields, the simulated surfaces separating two layers can be continuous and smooth. Borehole information is often incomplete in the sense that it does not provide direct information as to the exact layer some observed thickness belongs to. The latent Gaussian model proposed in this paper offers a natural solution to this problem by means of a Bayesian setting  with a Markov Chain Monte Carlo  (MCMC) algorithm that can explore all possible configurations compatible with the data. The model and the associated MCMC algorithm are validated on synthetic data and then applied to a subsoil in the Venetian Plain with a moderately dense network of cored boreholes.

\medskip


\end{abstract}

%%%%%%%%%%%%%%%%%%%%%%%%%%%%%%%%%%%%%%%%%%%%%%%%
%
\section{Introduction}
\label{sec:intro}
%%%%%%%%%%%%%%%%%%%%%%%%%%%%%%%%%%%%%%%%%%%%%%%%

The case study motivating this work is a subsoil  in the Venetian Plain with a moderately dense network of cored boreholes. Geologists and hydrogeologists managing this subsoil are in need of stochastic three-dimensional models of the stratigraphic sequence. The model should of course be conditioned to borehole data. The sequence of layers must correspond to the known regional stratigraphic sequence and, in addition, the surfaces separating the layers are required to be smooth and continuous.

Simulating a depositional (or stratigraphic) sequence conditionally on boreholes data has been and still is a long-standing problem in hydrogeology and in petroleum geostatistics. In the context of reservoir modeling, \cite{pyrcz2015stratigraphic} offers a comprehensive overview of the literature and a convincing conceptual framework in which methods are represented along a complexity gradient with one extreme corresponding to pixel based models with statistics and conditioning derived from the data and the other extreme representing geological concepts unconditional to local observations. As models tend to move away from the less complex extreme to the more complex one, they are less versatile and more difficult to condition \citep{pyrcz2015stratigraphic}. Easy-to-condition pixel based methods thus tend to be favored when data are dense, whereas rule-based or process-based models are preferred when conditioning data is sparse.  

Pixel based approaches, whether being based on variograms \citep{matheron1987conditional}, truncated Gaussian random fields and plurigaussian random fields \citep{beucher1993including,galli1994pros,armstrong2011plurigaussian,le2017modelling,le2018geostatistical}, transiograms \citep{carle1996transition}, MCP \citep{allard2011efficient,sartore2016spmc,benoit2018directional}, are well known and relatively easy to handle.  For these approaches, variogram and transiogram fitting is well understood and conditioning to well data is efficient, even for truncated Gaussian models \citep{marcotte2018gibbs}. However, one source of difficulty in the fitting procedure is the fact that  the processes and the amount of information are often anisotropic.  Typically for borehole data there is much more information along the depth than along horizontal directions. 

Multiple point statistics (MPS) approaches \citep{strebelle2002conditional,mariethoz2014multiple} require a training image when simulations are performed in two dimensions. Three-dimensional simulations are much more difficult to perform, since training cubes are rarely available at kilometer scales. Methods for combining images in three-dimensional simulations have been proposed \citep{comunian20123d,comunian2014training}.  But since a high degree of continuity is required for layers in this work, pixel based methods, including MPS, are not deemed appropriate. 

Object models such as Boolean models are more difficult to fit and to condition, in particular when accounting for non stationarity and erosion rules,  see for example \citet{syversveen1997conditioning} and \citet{allard2006conditional}. In addition, object models are not geologically appropriate for simulating sequences of layers.

Rule-based and process-based models incorporate some amount of understanding of the geological processes. They use rules to control the temporal sequence and spatial position of geological objects so as to 
mimic geological processes. Among others, they have been applied to fluvial systems, deepwater channel systems and turbiditic lobes systems. As particular cases of interest to this work are surface-based models. For simulating lobes in a turbidite reservoir, \cite{bertoncello2013conditioning} proposed a rule-based stacking of lobe-shape events with quite complicated sequential placement rules that depend partly on the already simulated events. The conditioning to well-log data and seismic data is achieved through sequential optimization. One of the limitation of this approach is that the variability between the conditional simulations is low, owing to the optimization approach. A second limitation recognized by the authors is that their method works best with limited amount of data.

This paper presents a new rule-based approach for simulating depositional sequences of surfaces conditionally to  lithofacies  thickness data. It is a stochastic model that  belongs to the \textit{Markov rules} sub-class of rule-based methods, see \cite{pyrcz2015stratigraphic} and appropriate references therein. The thickness of each layer is modeled by a transformed latent Gaussian random field allowing for null thickness.  The random fields are \textit{latent} because they can be unobserved on some parts of the domain under study, thanks to a truncation process.  Layers are sequentially stacked above each other following the regional stratigraphic sequence. By choosing adequately the variograms of these random fields, the simulated surfaces separating two layers can be continuous and smooth. Conditioning to the observed borehole data is made possible thanks to constrained Gaussian conditioning, as will be explained later on. 

A problem that has been barely addressed in geostatistical models for depositional sequences is the fact that borehole information is often incomplete in the sense that it does not provide direct information regarding the exact layers that have been observed. For example, let us consider that the stratigraphic sequence of the study domain contains several repetitions of a given lithofacies, say Clay. Consider also that the recorded data at one given borehole measures one single thickness for  Clay. A first possibility is that there is actually only one Clay layer at this location, but it could be any  of the several Clay layers of the regional stratigraphic sequence. Simulations should therefore account for this uncertainty. A second possibility is that the measurement actually corresponds to two (or more) Clay layers, one on top of the other, with missing intermediate layers at this location. In this case, the measured thickness should be shared between two layers.  The latent Gaussian model proposed in this paper offers a natural solution to this problem by means of a Bayesian setting with a  Markov Chain Monte Carlo (MCMC) algorithm that can explore all possible configurations compatible with the data. Notice that the approach proposed in \cite{bertoncello2013conditioning} does not address this problem at all.

The rest of this paper is organized as follows. Section \ref{sec:concept_model} is devoted to the conceptual model. In particular the difference between the (unique) regional stratigraphic sequence, referred to as the parent sequence, and the observed sequences is detailed. Section \ref{sec:stat_model} presents the stochastic model. In Section \ref{sec:MCMC} all details for Bayesian inference with an MCMC algorithm are given. It is then validated on a synthetic data set in Section \ref{sec:valid}. Finally, it is successfully applied to the Venetian Plain that motivated this work in Section 6. Some concluding remarks are then given in Section 7.

%%%%%%%%%%%%%%%%%%%%%%%%%%%%%%%%%%%%%%%%%%%%%%%%
%
\section{The conceptual model}
\label{sec:concept_model}
%%%%%%%%%%%%%%%%%%%%%%%%%%%%%%%%%%%%%%%%%%%%%%%%

\subsection{Notations}
Let us consider a spatial domain ${\mathcal S} \in \R^2$ and an interval $\mathcal{T} \subset \R^+$, which will correspond to ``depth''.  Note that depth can be converted into time through depositional processes, which is the reason why $t \in \mathcal{T}$ is used to denote depth. Let us also consider a family of $K$ lithofacies,  ${\mathcal C} = \{C_1,\dots,C_K\}$. The aim of this work is to build a process $X=\{X(s,t)\}$, defined at any point $(s,t) \in {\mathcal S} \times {\mathcal T}$ and taking values in $\mathcal C$. In words, at each location is associated one and only one lithofacies. The process must be continuous almost everywhere and the discontinuity surfaces should be smooth and they should have a general horizontal orientation. The process $X$ is observed along  depth at a finite number of locations $s_1,\dots,s_n$ and each observation corresponds to a drilled core, referred to as boreholes in the rest of this work. 

Let $X_i=\{X(s_i,t), t\in \mathcal{T}\}$ be one of these observations at site $s_i,\ i=1,\dots,n$, where $n$ is the number of sites. The observation $X_i$ is piece-wise constant, with $M_i$  discontinuities at different depths each time a new layer is encountered.
Therefore  the resulting information is a  sequence of facies and depths, referred to as the \textit{observed sequence}, $(\bC^o_i,\bT^o_i)$, where %(see Figure \ref{fig:facies}),
$\bC^o_i=(C^o_{1,i},\dots,C^o_{M_i,i})$ with $C^o_{j,i}\in\mathcal{C}$ for $j=1,\ldots,M_i$, and  $\bT^o_i=(T^o_{1,i},\dots,T^o_{M_i,i})$ with $T^o_{j,i} \in {\mathcal T}$ and $T^o_{1,i}<\cdots<T^o_{M_i,i}$. The depths are measured with respect to a  ground-level $T_{0,i}$. The thicknesses of each observed layers 
$\bZ^o_i=(Z^o_{1,i},\dots,Z^o_{M_i,i})$ can be derived from the depths, with  $Z^o_{j,i} = T^o_{j,i} - T^o_{j-1,i}$, $j=1,\ldots,M_i$. Finally, the last layer is assumed to be completely observed, that is the depth $Z^o_{M_i,i}$ is assumed to be not censored.

\subsection{Parent sequence}
The working hypothesis is that there exists a common lithological sequence of facies hereafter referred to as the ``parent sequence'', which is compatible with all observed sequences in the area of study in the sense that each observed sequence can be obtained from the parent sequence by deleting some layers of the parent sequence.

This sequence  can result from the prior knowledge of the scientists.
Alternatively, it can be   derived from the observed data. From a mathematical viewpoint, there always exists a parent sequence. For example, it can easily be obtained by simply stacking all observed sequences into a single sequence. Then, each observed sequence of layers is simply obtained by ``deleting'' all other observed sequences. Obviously, this parent sequence is of no modeling interest, but it is mathematically important since it provides a proof of the existence of this concept. In general, very long parent sequences are uninteresting from a modeling point of view. In accordance with a parsimony principle, one should seek the shortest possible parent sequences. Clearly, there is only a finite number of parent sequences of minimal length. Such parent sequences could be built using discrete optimization algorithms, or it could be provided by scientists, based on prior geological knowledge. Either way, how minimal parent sequences are obtained is a subject out of the scope of the present research, and this route is not pursued any longer. 

From now on  it will be considered that the parent sequence is known, and that it is one of the minimal length parent sequences.  The parent sequence of length $M$ will be denoted $\bC = (C_1,\dots,C_M)$, $C_i\in\mathcal{C}$, with  $M\ge \max\{ K, M_1,\dots,M_{n}\}$. 

\subsection{From the parent sequence to the observed sequences}
\begin{figure}
    \centering
    \includegraphics[width=9cm]{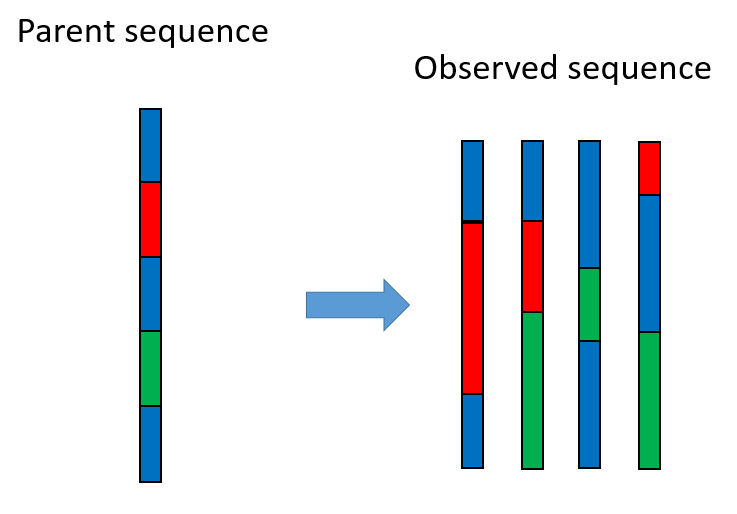}
    \caption{Parent sequence and four possible incomplete observed sequences. Since the parent sequence is conceptual, thicknesses are only meaningful in the observed sequences.}
    \label{fig:parentSeq}
\end{figure}

When analyzing sequences of lithofacies, it is quite common that some facies are unobserved at one or several boreholes. In order to allow for this, each observed sequence  at each  site $s_i$ is therefore a subset of a complete sequence $(\bC,\bT_i)$  corresponding to the parent sequence. The corresponding vector of complete thickness is $\bZ_i$, and in contrast to the observed ones, some  thickness  $Z_{j,i} = T_{j,i} - T_{j-1,i}$, $j=1,\ldots,M$ can be equal to zero. In this case, the corresponding layer is unobserved at location $s_i$. When $M_i < M$, the sequence at $s_i$ is an \textit{ incomplete sequence}, and $\bC^o_i$ is a sub-sequence of $\bC$. The complete data will be denoted $\bX=\{(\bC,\bZ_i),\, i=1,\ldots,n \}$ and $\bX^o=\{(\bC^o_i,\bZ^o_i),\, i=1,\ldots,n \}$ will denote the observed data. In the following, $O(\cdot)$ will denote the mapping such that $\bX^o=O(\bX)$. Figure \ref{fig:parentSeq} illustrates a parent sequence and 4 different possible observed sequences.

%%%%%%%%%%%%%%%%%%%%%%%%%%%%%%%%%%%%%%%%%%%%%%%%
%
\section{Statistical setting}
\label{sec:stat_model}
%%%%%%%%%%%%%%%%%%%%%%%%%%%%%%%%%%%%%%%%%%%%%%%

\subsection{Stochastic model}
\label{sec:stoch_mode}
The stochastic model requires a univariate model for the marginal distribution of the thicknesses and a spatial model to account for the lateral continuity of the layers. Thicknesses  are modeled using  positive zero inflated random variables in order to account for the many 0s resulting from incomplete observed sequences. Among many possible models, latent truncated Gaussian models \citep{allcroft2003latent,baxevani2015spatiotemporal,benoit2018stochastic}, also referred to as Tobit models \citep{liu2019statistical} in econometrics, are flexible models that allow easily a geostatistical modeling.  Spatial dependence among the thicknesses belonging to a same layer is introduced by means of a truncated  Gaussian random field.  More precisely, for $j=1,\dots,M$, let  $W_j(s), s \in \mathcal{S}$ be a standardized  Gaussian random field that, for simplicity,  will be supposed stationary  with covariance function $\mbox{cov}[W_j(s),W_j(s')] = \rho_j(s-s';\xi_j)$, where $\rho_j$ is a parametric correlation function and $\xi_j$ the vector of associated parameters. The thickness field $\{Z_j(s),s\in \mathcal{S}\}$ is defined as:
\begin{equation}
Z_j(s) = \varphi_j \bigl(W_j(s)-\tau_j\bigr)\quad  \hbox{if} \quad W_j(s) > \tau_j,
\label{eq:TruncRF}
\end{equation}
and $Z_j(s) = 0$ otherwise, where $\tau_j$ is a threshold and $\varphi_j(\cdot)$ is a  continuous one-to-one mapping from $\R_+$ to $\R_+$. The probability of positive thickness $\Pr(Z_j(s) > 0)$ will be denoted by $p_j$. With this construction, null thickness has a positive probability, since $\Pr(Z_j(s) = 0)=1-p_j = \Phi(\tau_j)>0$, where $\Phi(\cdot)$ is the cumulative probability function of the standard Gaussian random variable. Parameters of the stochastic model can be expressed equivalently in terms of $\tau_j$ or $p_j$ and in the sequel the second setting is chosen.
One particular case that will be used later is to set $\varphi_j(x) = \mu_j x^{\beta_j}, x>0$ with $\beta_j, \mu_j> 0$.  When $\beta_j=1$, one gets
\begin{equation}
E\bigl[Z_j(s)\bigr] = \mu_j \left( \frac{\phi(\tau_j)}{1-\Phi(\tau_j)} - \tau_j \right)\quad 
Var[ Z_j(s)] = \mu_j^2\left[1+ \frac{\phi(\tau_j)}{1-\Phi(\tau_j)}\left(\tau_j -  \frac{\phi(\tau_j)}{1-\Phi(\tau_j)}\right)\right],
\label{eq:exp_thickness}
\end{equation}
where $\phi(\cdot)$ is the density function of the standard Gaussian random variable. When $\beta_j$ is not an integer, the moments of $Z_j(s)$ involve hypergeometric functions and are not reported here. From Eq. \eqref{eq:exp_thickness} it is clear that the expectation and standard deviation of the thickness of layer $j$ are both proportional to the parameter $\mu_j$. The covariance function $\rho_j$ must be smooth enough in order to generate regular thicknesses. For example, choosing that $\rho_j$ is twice differentiable at the origin leads to a mean-squared differentiable random field $W_j$ and, as a consequence, to a mean-squared differentiable random field for the thicknesses since $\varphi_j$ is continuous and entails locally finite boundaries of the non null thickness sets. The depth surfaces $\{T_j(s),s\in \mathcal{S}\}$ are then obtained by adding up the thickness fields. Starting from a fixed and known ground-floor $T_0=\{T_0(s), s \in \mathcal{S}\}$ one sets 

\begin{equation*}
T_j(s) = T_{j-1}(s) + Z_j(s) = T_0(s)+\sum_{i=1}^j Z_i(s),\qquad j=1,\dots,M.
\end{equation*}
Finally, the random fields $W_j$ are assumed to be independent, since they relate to independent depositional processes.

\subsection{Complete likelihood} 
Since layers are assumed to be independent, the complete likelihood factorizes into a product of $M$ likelihoods
\begin{equation}
L(\theta;\bX)=\prod_{j=1}^{M} L_j(\theta_j;{Z}_{j,1},\ldots,{Z}_{j,n}),
\label{eq:CompleteLik}
\end{equation}
where  $\theta_j=(p_j,\mu_j,\beta_j,\xi_j), j=1,\dots,M$ and $\theta = (\theta_1,\dots,\theta_M)$.

In the sequel $\phi_{k}(\cdot,\boldsymbol{\mu},\boldsymbol{\Sigma})$ and $\Phi_{k}(\cdot,\boldsymbol{\mu},\boldsymbol{\Sigma})$ denote   the  density and the cumulative distribution function of a $k-$multivariate Gaussian random variable with  mean vector $\boldsymbol{\mu}$ and covariance matrix $\boldsymbol{\Sigma}$. 
Let us consider now a layer $j \in \{1,\dots,M\}$. For convenience, thicknesses and the corresponding locations are reordered such that the first $n_j$  thicknesses $Z_{j,1},\ldots,Z_{j,n_j}$  correspond to the  positive values and the remaining $\ell_j=n-n_j$ ones are 0. The complete-data likelihood of the single layer $j$ is
\begin{equation}
L_j(\theta_j;{Z}_{j,1},\ldots,{Z}_{j,n}) =
f_j({Z}_{j,1},\ldots,{Z}_{j,n_j};\theta_j)F_j(0,\ldots,0,|{Z}_{j,1},\ldots,{Z}_{j,n_j};\theta_j).
\label{eq:CompLik_j}
\end{equation}
The density $f_j({Z}_{j,1},\ldots,{Z}_{j,n_j};\theta_j)$ is given by
\begin{equation}
    f({Z}_{j,1},\ldots,{Z}_{j,n_j};\theta)=\phi_{n_j}({W}_{j,1},\ldots,{W}_{j,n_j};\mathbf{0},\mathbf{\Sigma}_j) \prod_{i=1}^{n_j} J_{\varphi_j^{-1}}(Z_{j,i}).
    \label{eq:density_j}
\end{equation}
where $\boldsymbol{\Sigma}_j=\boldsymbol{\Sigma}_{n_j,n_j}=[\rho(s_i-s_k;\xi_j)]_{i,k=1,\ldots,n_j}$,
$W_{j,i} = \varphi_j^{-1}\bigl( Z_{j,i} \bigr) +\tau_j,\quad i=1,\ldots,n_j,$
and  $J_{\varphi_j^{-1}}(Z_{j,i})$ is the Jacobian of ${\varphi_j^{-1}}$ computed at $Z_{j,i}$.
The conditional probability $F_j(0,\ldots,0|{Z}_{j,1},\ldots,{Z}_{j,n_j};\theta)$
is given by
\begin{equation}
F_j(0,\ldots,0|{Z}_{j,1},\ldots,{Z}_{j,n_j};\theta)=\Phi_{l_j}(\tau_j,\ldots,\tau_j;\mathbf{m}_j,\mathbf{V}_j)
\label{eq:conditional_j}
\end{equation}
where the mean vector $\mathbf{m}_j$ and covariance matrix $\mathbf{V}_j$ can be easily derived using the kriging equations \citep{Cressie:1993,Chiles2012}:
\begin{equation}\label{eq:kriging}
\mathbf{m}_j = \mathbf{\Sigma}_{\ell_j,n_j} \mathbf{\Sigma}^{-1}_{n_j,n_j} \mathbf{W}_{n_j};\quad
\mathbf{V}_j = \mathbf{\Sigma}_{\ell_j,\ell_j} -  \mathbf{\Sigma}_{\ell_j,n_j} \mathbf{\Sigma}^{-1}_{n_j,n_j}  \mathbf{\Sigma}_{n_j,\ell_j},
\end{equation}
with $\mathbf{W}_{n_j}=({W}_{j,1},\ldots,{W}_{j,n_j})^\prime$ and the matrices $\mathbf{\Sigma}_{\ell_j,n_j}$ and $\mathbf{\Sigma}_{\ell_j,\ell_j}$ being defined in similar ways as $\mathbf{\Sigma}_{n_j,n_j}$. To summarize, the complete data likelihood in \eqref{eq:CompleteLik} becomes:
\begin{eqnarray}
	L(\theta;\bX)&=&\prod_{j=1}^{M} L_j(\theta;{Z}_{j,1},\ldots,{Z}_{j,n}) \nonumber\\
	&=& \prod_{j=1}^{M}\phi_{n_j}({W}_{j,1},\ldots,{W}_{j,n};\mathbf{0},\mathbf{\Sigma}_j)\  \prod_{i=1}^{n_j} J_{\varphi_j^{-1}}(Z_{j,i})\Phi_{l_j}(\tau_j,\ldots,\tau_j;\mathbf{m}_j,\mathbf{V_j}).
	\label{eq:CompleteLik2}
\end{eqnarray}
In the particular case $\varphi_j(x) = \mu_j x^{\beta_j}$ that will be considered below, the Jacobian simplifies to 
\begin{equation}
J_{\varphi_j^{-1}}(Z_{j,i}) = \frac{1}{\mu_j \beta_j}  \left( \frac{Z_{j,i}}{\mu_j}  \right)^{1-1/\beta_j}.
\label{eq:jacobian}
\end{equation}

\subsection{Observed Likelihood}
In principle the observed likelihood is related to the complete likelihood through
\begin{equation}
L(\theta;\bX^o)=\int_{\{\bX:\bX^o=O(\bX)\}} L_X(\theta;\bX) d\bX.
\label{eq:obs_lik}
\end{equation}
However even for moderately long parent sequence and number of 0 thicknesses, the space $\{\bX:\bX^o=O(\bX)\}$ is difficult to explore and the integral \eqref{eq:obs_lik} becomes intractable. These difficulties are illustrated with two examples. At some site, let us consider an observed sequence $(\bC^o,\bT^o)$ and  the corresponding  thicknesses $\bZ^o$. Here, the reference to the site is dropped for the sake of clearer notations. Recall that since the sequence $\bC^o$ must be compatible with the parent sequence $\bC$,  $\bC^o$ is obtained by deleting some layers of $\bC$.  

Table \ref{tab:ex1} shows an example of a parent sequence $\bC$ with three categories: \texttt{Blue}, \texttt{Red}  and \texttt{Green}. The observed sequence $\bC^o$ is incomplete. Several augmented sequences $\bC^a$ with corresponding  depths $\bT^a$ are possible. Since in the observed series the first \texttt{Blue} is followed by \texttt{Red}, the sub-sequence [\texttt{Blue}-\texttt{Red}] must correspond to the beginning of the parent sequence. Regarding the second occurrence of \texttt{Blue},  three cases can be distinguished: i) it corresponds only to the third layer of $\bC$ with 4th and 5th layers having null thickness; ii) it corresponds only to the fifth layer, in which case the 3rd and 4th layers have null thickness; iii) it corresponds partly to the 3rd and partly to the 5th layers. Then, only the 4th layer has 0 thickness. In this last case, an intermediate,  latent, transition at depth $\tilde{T}$ with $T^o_2 \leq \tilde{T} \leq T^o_3$ must be introduced. These augmented series are all possible, but some will be more likely than others, depending on the parameters of the model. In Appendix A an even more complex example is provided. Only some of the possible configurations are shown. They are too numerous and complex to be completely listed, even for short parent sequences.

In order to estimate the parameters of the model a data augmentation algorithm \citep[Ch. 5]{Tanner1996} can be exploited  where the complete sequences that are compatible with the observed ones are explored. A Bayesian approach will be adopted for the inference of the parameters and a  Markov Chain Monte Carlo (MCMC) algorithm will be designed in Sect. \ref{sec:MCMC}. But first, simulation when all parameters are known and when all sequences are complete is shown.

\begin{table}[ht]
	\caption{Example of a parent sequence $\bC$	with an observed sequence $\bC^o$ and  several possible augmented sequences with corresponding transition depths  and thicknesses \label{tab:ex1}}
	\begin{center}
		\begin{tabular}{ccc|ccc|ccc|ccc}
			\hline\hline
			Parent & \multicolumn{2}{c|}{Observed} & \multicolumn{9}{c}{Possible augmented sequences}\\
			\hline
			$\bC$ & $\bC^o$ & $\bT^o$ &$\bC^a$ & $\bT^a$  & $\bZ^a$ &$\bC^a$ & $\bT^a$  & $\bZ^a$ &$\bC^a$
			& $\bT^a$  & $\bZ^a$ \\
			\hline\hline 
			\texttt{Blue} & \texttt{Blue}  & $T^o_1$ & \texttt{Blue} & $T^o_1$ & $T^o_1$     & \texttt{Blue} & $T^o_1$ & $T^o_1$     & \texttt{Blue}  & $T^o_1$ & $T^o_1$ \\ 
			\texttt{Red} & \texttt{Red}  & $T^o_2$ & \texttt{Red} & $T^o_2$ & $T^o_2-T^o_1$ & \texttt{Red} & $T^o_2$ & $T^o_2-T^o_1$ & \texttt{Red}  & $T^o_2$ & $T^o_2-T^o_1$\\
			\texttt{Blue} & \texttt{Blue}  & $T^o_3$ & \texttt{Blue} & $T^o_3$ & $T^o_3-T^o_2$ &   & $T^o_2$ & 0         & \texttt{Red}  & $\tilde{T}$ & $\tilde{T}-T^o_2$\\
			\texttt{Green} & -- & --    &   & $T^o_3$ & 0          &   & $T^o_2$ & 0         &   & $\tilde{T}$ & 0 \\
			\texttt{Blue} & -- & --    &   & $T^o_3$ & 0          & \texttt{Blue} & $T^o_3$ & $T^o_3-T^o_2$ & \texttt{Blue} & $T^o_3$ & $T^o_3-\tilde{T}$\\
			\hline\hline
		\end{tabular}
	\end{center}
\end{table}

\subsection{Simulation}
\label{sec:simulation}
\begin{figure}
	\begin{center}
\begin{tabular}{ccc}
\includegraphics[height=4cm,width=1.1cm]{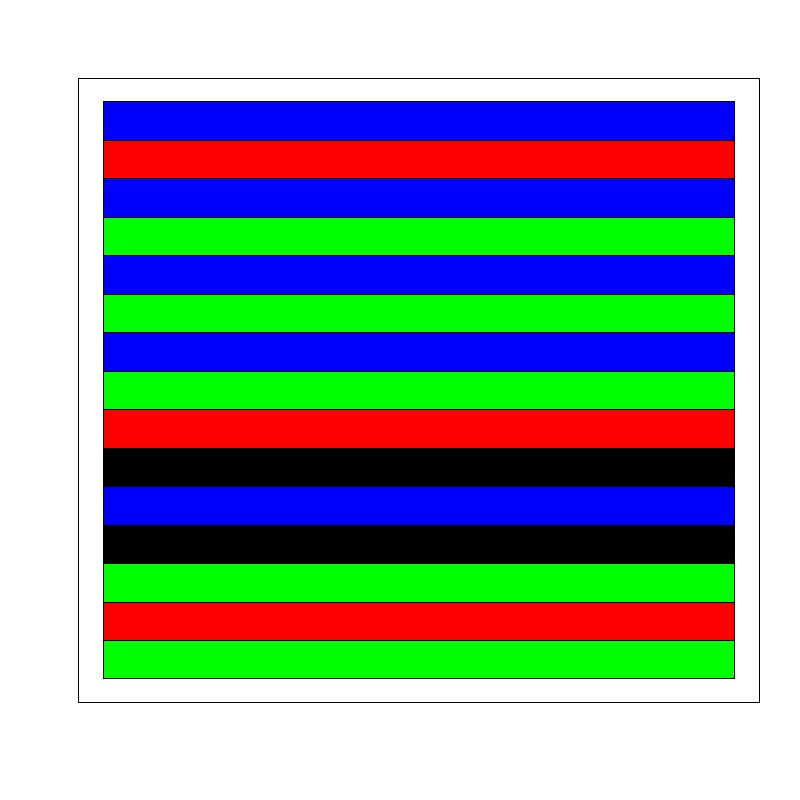}
     &  \includegraphics[width=10.1cm,height=4cm]{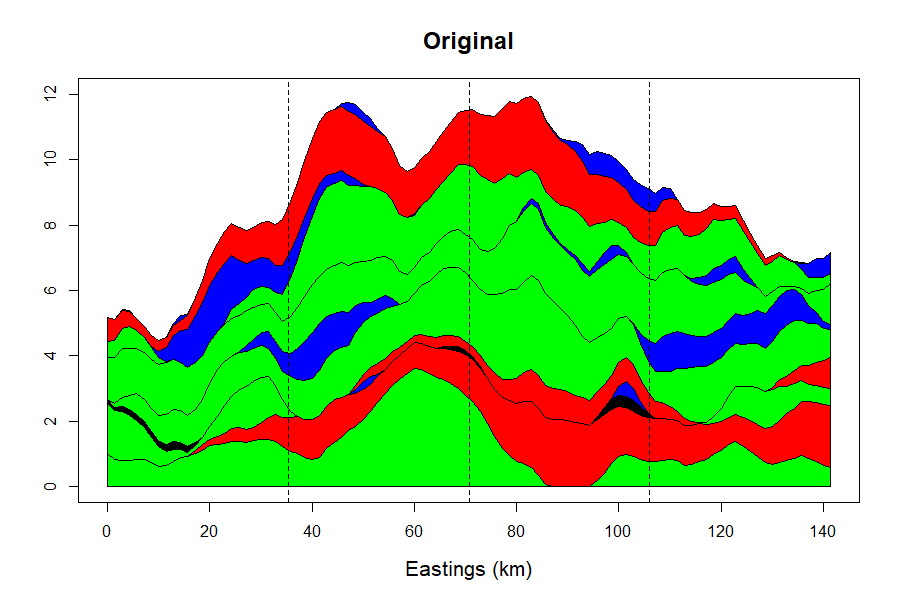} &  \includegraphics[height=4cm]{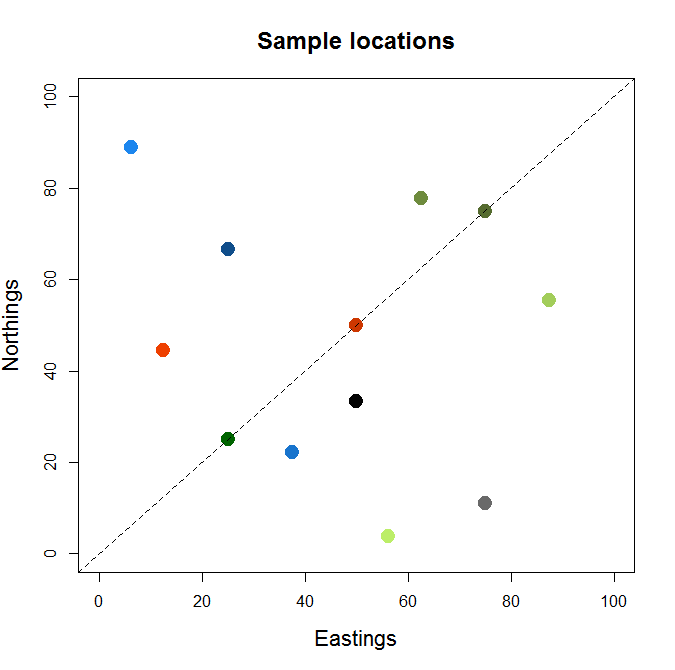}\\
(a) & (b) & (c)
\end{tabular}	
		
	\end{center}
	\caption{Simulation experiment: (a) Parent sequence  of length 15, with 4 lithofacies \{\texttt{Black}-\texttt{Red}-\texttt{Blue}-\texttt{Green}\};	(b) Cross-section of a two-dimension simulation along the diagonal of ${\mathcal S} =[0,100] \times [0,100]$. See Table \ref{tab:parameters} for the  parameters; (c) Locations of the twelve boreholes.}\label{fig:sim}
\end{figure}

Unconditional simulation is straightforward when the transformation $\varphi_j$ and the parameters $\theta_j$, $j=1,\ldots,M$, are known. All that is required is to simulate $M$ random fields $W_j, j=1\dots,M$ and then to apply \eqref{eq:TruncRF} in order to transform the Gaussian process into a thickness surface. Figure \ref{fig:sim} illustrates a cross-section of a two-dimension simulation over ${\mathcal S} =[0,100] \times [0,100]$ with four lithofacies \{\texttt{Black}-\texttt{Red}-\texttt{Blue}-\texttt{Green}\} and $\varphi_j(x) = \mu_j x$, that is $\beta_j=1$ for all categories.
The parent sequence has 15 layers (see Fig. \ref{fig:sim}-(a))  and  stochastic models for layers with  the same lithofacies have identical set of parameters.  Thicknesses have been simulated  using Gaussian random fields with a 
Mat\'ern covariance function 
\begin{equation}
\rho(h;\nu,\alpha,\sigma^2) =  \frac{\sigma^2}{2^{\nu-1}\Gamma(\nu)} \left( \frac{||h||}{\alpha}\right)^\nu K_\nu \left( \frac{||h||}{\alpha}\right),  \quad h \in \R^2,
\label{eq:matern}
\end{equation}
where $\nu>0$ is a smoothness parameter, $\alpha>0$ a range parameter and $\sigma^2$ the sill. $\Gamma$ is the gamma function and $K_\nu$ is the modified Bessel function of the second kind of order $\nu$. Here, the smoothness parameter has been set to $\nu=3/2$ and $\sigma^2=1$, which leads to the simplified expression  $\rho_j(h;\alpha_j) = (1+||h||/\alpha_j)\exp(-||h||/\alpha_j)$, where $\alpha_j$ is a range parameter.  The set of the parameters in the simulation experiment is shown in Table \ref{tab:parameters}. 
\begin{table}
\caption{Parameters for the simulation example.}
\label{tab:parameters}
    \centering
		\begin{tabular}{ccccc}
	\hline\hline
	& \texttt{Black} & \texttt{Red} & \texttt{Blue} & \texttt{Green} \\
	\hline
	$\mu, \beta$ & 1 & 1& 1& 1\\
	$p$   & 0.3   & 0.8 & 0.3 & 0.8 \\
	$\alpha$  & 20    & 20  & 10  & 10\\
	\hline\hline
\end{tabular}

\end{table}{}

Twelve synthetic {boreholes} have been located in ${\mathcal S}$. Three of them are placed along the diagonal at coordinates $(25,25)$, $(50,50)$ and $(75,75)$. Nine others are randomly located (see Fig. \ref{fig:sim}-(c)). For each category, the observed frequencies along these twelve boreholes  are $(0.58,0.83,0.28,0.80)$. Notice that \texttt{Black} is therefore highly over-represented. The average thicknesses computed along the {boreholes} are $(0.31,1.31,0.62,1.25)$ for each of the four categories, whilst the theoretical expectations of each category computed as per \eqref{eq:exp_thickness} are respectively $(1.8,1.1,1.8,1.1)$. Note here that \texttt{Black} and \texttt{Blue} are very unlikely to be directly stacked above each other, while it is often the case for \texttt{Red} and \texttt{Green}.

Conditional simulation is relatively easy to implement when the parameters are known and when complete sequences of thicknesses are available, including all null thicknesses. Care must be taken when simulating  values from the Gaussian distribution that are below the thresholds $\tau_j$, but otherwise the algorithm, shown in Algorithm 1, is rather straightforward. Simulations of the truncated Gaussian values are done by calling the function \texttt{rmvnorm} of the \texttt{R} package \texttt{mvtnorm} \citep{GenzPackage}. The reader is referred to \cite{Chiles2012} for a general exposition on unconditional simulations and conditional simulations using Kriging techniques.

\begin{algorithm}[thb]
\caption{Conditional simulation when all sequences and all parameters are known} 
\begin{algorithmic}[1]
\REQUIRE Data with complete sequences; transform functions $\varphi_j, j=1,\dots,M$
\REQUIRE All parameters
\FOR{$j=1$ to $M$}
\STATE Compute the vector $\mathbf{W}_{n_j}=(W_{j,1},\ldots,W_{j,n_j})$ where $W_{j,k} = \varphi_j^{-1}(Z_{j,k})$ corresponding to $Z_{j,k}>0, k=1,\dots,n_j$ 
\STATE Compute $\mathbf{m}_j$ and $\mathbf{V}_j$ according to \eqref{eq:kriging}
\STATE Draw a  vector of length $\ell_j$ from a truncated multivariate Gaussian distribution, \\        $\mathbf{W}_{l_j} \sim {\mathcal{TN}}_{\ell_j}(\mathbf{m}_j,\mathbf{V}_j;-\infty, \tau_j)$, for which each component must be below $\tau_j$.
\STATE Set $\mathbf{W}_j = (\mathbf{W}_{n_j},\mathbf{W}_{\ell_j})$
\STATE Simulate a Gaussian random field $F_j$ conditionally   on $\mathbf{W}_j$ 
\STATE Transform the field $F_j$ into the thicknesses according to \eqref{eq:TruncRF}
\ENDFOR
\end{algorithmic}
\end{algorithm}

\clearpage

%%%%%%%%%%%%%%%%%%%%%%%%%%%%%%%%%%%%%%%%%%%%%%%%
%
\section{Bayesian inference with a  Markov Chain Monte Carlo algorithm}
\label{sec:MCMC}
%%%%%%%%%%%%%%%%%%%%%%%%%%%%%%%%%%%%%%%%%%%%%%%

\subsection{Sampling all possible configurations}
\begin{figure}
    \centering
    \includegraphics[width=14cm]{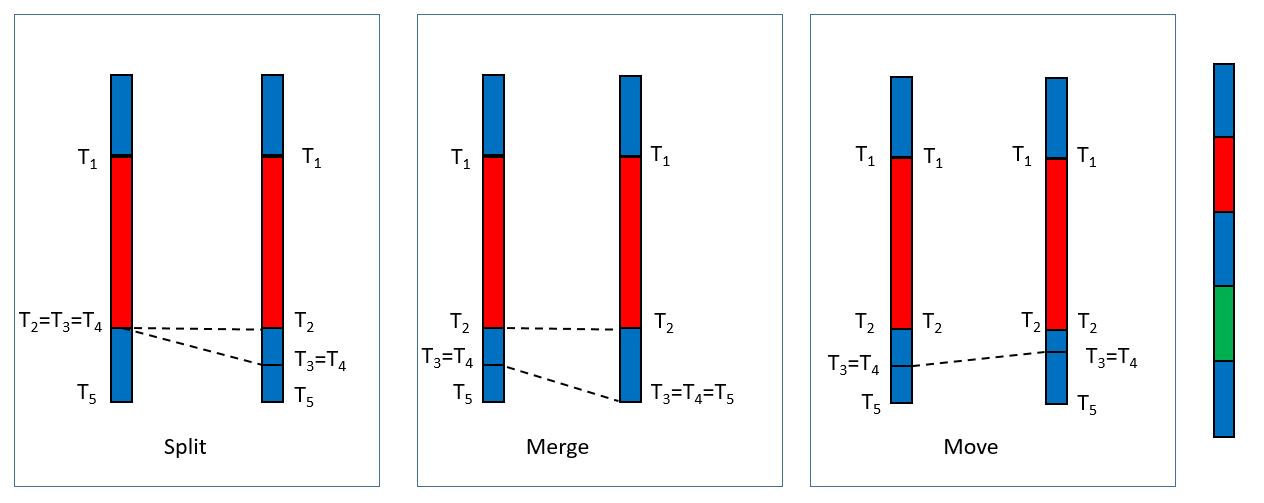}
    \caption{Elementary moves in an incomplete observed sequence. Note that the layer \texttt{Green} is unobserved. From left to right: \textit{Split}, \textit{Merge} and \textit{Displace}.}
    \label{fig:moves}
\end{figure}

In order to sample within all possible configurations of the augmented sequence at a given site $s_i$ that are compatible with the parent sequence, the   Markov Chain Monte Carlo (MCMC) algorithm must be able to delete a layer, to add a new layer or to displace the limit between two layers of the same category. Recall that the limit between two different categories are hard conditioning data that cannot be changed. These elementary moves, illustrated in Fig. \ref{fig:moves}, are now detailed.
\begin{description}
\item \textit{Split}: A state is split into two successive states of the same category. A split is only possible if it is compatible within the parent sequence. For example, in Fig. \ref{fig:moves}, the \texttt{Blue} layer at the bottom can be split into two layers since the parent sequence contains a second \texttt{Blue} layer. In Table \ref{tab:ex2} the situation in panel number 4 can be obtained by splitting the state \texttt{Red}, either in panel 2 or in panel 3.
When a state is split, a new transition depth, denoted  $t_i$ in Table \ref{tab:ex2}, must be introduced. The thickness is split in two thicknesses accordingly.
\item \textit{Merge}: This move is the opposite move of \textit{Split}. Two successive states in the same category are merged together. The corresponding depth is removed and the resulting thickness is the sum of the two merged thicknesses.  
\item \textit{Displace}: Here, the augmented sequence is not changed, but the intermediate value between two successive states of the same category is changed. The corresponding thicknesses are then updated.     
\end{description}

{}

It is easy to verify that starting from any initial configuration compatible with the parent sequence, any other configuration can be reached  by combining finite numbers of \textit{Split}, \textit{Merge} and \textit{Displace}. Hence, if these moves are used as building blocks of a MCMC algorithm, the resulting Markov Chain will be ergodic. At each {borehole}, one of the three moves is proposed with probabilities $(p_S,p_{M},p_{D})$ with  $p_S+p_{M}+p_{D}=1$. If the move is possible, it is accepted according to  Metropolis-Hasting acceptance ratio described in Sect. \ref{sec:general_algo}.

\subsection{Choosing the priors}
\label{sec:priors}

Priors must be defined for all parameters of the model. For the parameters of the transform functions $\varphi_j$, $1-p_j=\Phi(\tau_j)$ and $\beta_j$,  uninformative flat priors have been chosen on the intervals $(0,1)$ and $(0.25,4)$ respectively.
Regarding the covariance function, the Mat\'ern covariance function in \eqref{eq:matern}
has been chosen for its great flexibility thanks to three parameters: $\xi = (\nu,\alpha,\sigma)$, for smoothness, range and sill, respectively.

However, it is  known that the joint estimation of these parameters is difficult in a Bayesian context, in particular if the number of data is small. \cite{zhang2004inconsistent}  showed that for a Mat\'ern covariance function the only quantity that can be estimated consistently under in-fill asymptotics is $\sigma^2 \alpha^{-2\nu}$.
As a consequence, since the parameter $\mu^2$ behaves as the marginal variance of the random field, using uninformative flat priors for $(\mu,\alpha,\nu)$ is expected to provide poor posterior distributions for these parameters. This was indeed confirmed on preliminary MCMC runs (results not reported here).  It was thus decided to fix the smoothness parameter $\nu$ among the values $(1/2,3/2,5/2)$ that would provide the highest likelihood. The above values correspond to covariance functions being the product of an exponential and a polynomial of order $p$ with $p=0,1,2$ respectively, namely $\rho(r;1/2,\alpha,1) = \exp(-r/\alpha)$,  $\rho(r;3/2,\alpha,1) = (1+ r/\alpha)\exp(-r/\alpha)$ and $\rho(r;5/2,\alpha,1) = (1+ r/\alpha + r^2/(3\alpha^2))\exp(-r/\alpha)$. 

\cite{simpson2017penalising}  proposed an approach for building priors that are based on penalizing the complexity to a base model. For example, a random effect with positive variance is an extension (a more complex version) of random effect with null variance. Similarly, a random field with a finite range is an extension (a more complex version) of a random field with infinite range. Indeed, if the range is infinite, the random field is perfectly correlated and its spatial variance is null. Penalized Complexity (PC) priors are then defined as the only priors that: i) use the Kullback-Leibler divergence as a measure between the extended and the base models; ii) have a penalization that increases with the distance at a constant rate.

\cite{fuglstad2019constructing} derived the PC priors for a Mat\'ern covariance with parameters $\sigma$, $\alpha$ and $\nu$, when $\nu$ is fixed. They showed that the joint PC prior corresponding to a base model with infinite range and zero variance when $d=2$ is
\begin{equation}
    \pi(\sigma,\alpha) = \lambda_{\alpha} \alpha^{-2} \exp\bigl( -\lambda_{\alpha}/ \alpha\bigr) \lambda_{\sigma} \exp\bigl( -\lambda_{\sigma} \sigma\bigr),
    \label{eq:PCpriors}
\end{equation}
where $\lambda_{\alpha} = -\ln (\epsilon_\alpha) \alpha_0$ and $\lambda_{\sigma} = -\ln (\epsilon_\sigma) /\sigma_0$, and the values of $\lambda_{\alpha}$ and $\lambda_{\sigma}$ are such that 
$P(\alpha < \alpha_0) = \epsilon_\alpha$ and $P(\sigma > \sigma_0) = \epsilon_\sigma$. By choosing small probabilities $\epsilon_\alpha$ and $\epsilon_\sigma$, the range is lower-bounded above $\alpha_0$ and the standard deviation is upper bounded at $\sigma_0$ with probability $1-\epsilon_\alpha$ and $1-\epsilon_\sigma$, respectively. PC priors described in \eqref{eq:PCpriors} will be used throughout, where $\mu$ plays the role of the standard deviation as shown in Eq. \eqref{eq:exp_thickness} in Sect. \ref{sec:stoch_mode}. 

\subsection{General description of the algorithm}
\label{sec:general_algo}

Each parameter in each category is updated iteratively in a Metropolis-within-Gibbs algorithm \citep{gelfand2000gibbs}. A new value is proposed according to symmetric transition kernels, for which it is equally likely to move from a current value $y^c$ to a new value $y^n$ than the opposite. Let us denote $\theta^c$ and $\theta^n$ respectively the current and the proposed vector of parameters $\theta$. Let us further denote  $\pi(\cdot)$ the prior density of $\theta$. The acceptance ratio is then
\begin{equation}
A(\theta^c,\theta^n) = \frac{L(\theta^{n}; \bX) \pi(\theta^n)}{L(\theta^{c}; \bX) \pi(\theta^c)}.
\label{eq:ar_theta}
\end{equation}
When sampling the configurations thanks to one of the possible moves \textit{Split}, \textit{Merge} and \textit{Displace}, a new configuration $\bX^n$ is proposed, $\bX^c$ being the current one. In this case the acceptance ratio is
\begin{equation}
A(\bX^c,\bX^n) = \frac{L(\theta; \bX^n)}{L(\theta ; \bX^c)}.
\label{eq:ar_config}
\end{equation}
The proposals are accepted if the acceptance ratios $A(\cdot,\cdot)$  are larger than one. Otherwise, they are accepted with a probability equal to the ratio. The proposal in the Metropolis-Hasting step are random walk proposals aiming at an acceptance rate above $0.5$. For sampling new configurations at each {borehole} in turn, a possible move is drawn according to the probabilities $p_S=p_M=p_D=1/3$. Then, it is checked whether such move is feasible within this borehole. If several moves are possible, one is selected uniformly among all possible moves in that borehole, and a new configuration is proposed. The whole procedure is summarized in Algorithm 2.
  
\begin{algorithm}[htb]
\caption{MCMC procedure} 
\begin{algorithmic}[1]
\REQUIRE Data; parent sequence; transform functions $\varphi_j, j=1,\dots,M$
\REQUIRE Initial values and priors for all parameters
\REQUIRE Number of iterations, $N$
\FOR{$i=1$ to $N$} 
\FOR{ each parameter $ \eta \in \{p, \mu, \beta, \alpha\}$}
\FOR{$j=1$ to $M$} 
\STATE Propose new $\eta_j$ according to transition kernel
\STATE Compute acceptance ratio, $A$  using \eqref{eq:ar_theta}
\STATE Generate $U \sim {\mathcal U}[0,1]$; accept new $\eta_j$ if ($U \leq A)$
\ENDFOR
\ENDFOR
\FOR{ Borehole $k=1$ to $n$}
\STATE Draw a move $\in \{Split, Merge, Displace\}$ according to the probabilities $(p_S,p_M,p_D)$
\STATE Check for feasibility within {borehole} $k$
\IF{(move is feasible)}
\STATE Draw uniformly one among all possible moves
\STATE Compute acceptance ratio, $A$ using \eqref{eq:ar_config}
\STATE Generate $U \sim {\mathcal U}[0,1]$; accept the move if $(U \leq A)$
\ENDIF
\ENDFOR
\ENDFOR
\end{algorithmic}
\end{algorithm}

\section{A synthetic data example}
\label{sec:valid}

The MCMC algorithm described above is first validated on the synthetic data-set described in Sect. \ref{sec:simulation} and illustrated in Fig. \ref{fig:sim}. It was coded in {\tt R} using standard functions and our own code for the \textit{Split}, \textit{Merge} and \textit{Displace} movements. Most of the running time is spent in computing the simultaneous probabilities of being below $0$ in \eqref{eq:conditional_j}. This is done by calling the function {\tt pmvnorm} of the {\tt R} package \textit{mvtnorm} \citep{GenzPackage,GenzBretz2009}. Uniform priors are used for the parameters $p_j$ and $\beta_j$, respectively on $(0,1)$ and $(.25,4)$, while PC priors are used for the parameters $\mu_j$ and $\alpha_j$, as described in details in Sect. \ref{sec:priors}. Here, the setting was $\epsilon_\alpha = \epsilon_\mu=0.01$, with $\alpha_0=3$ and $\mu_0=10$. Algorithm 2 is run for $30,000$ iterations, after a burn-in period of $2,500$  iterations. Values of parameters are then sampled every $50$ iterations. The proposals in the Metropolis-Hasting steps follow a uniform random walk with increments in $[-0.4,0.4]$ for $\mu_j$ and $\beta_j$, in $[-0.15,0.15]$ for $p_j$ and in $[-3,3]$  for the range $\alpha_j$. With these choices, the observed acceptance ratio lies between $0.43$ and $0.57$, depending on the parameters. This dataset being quite constrained, the acceptation ratio for exploring new configurations is only $6.78\, 10^{-5}$.

\subsection{Estimation of the parameters}

\begin{figure}
    \centering
    \includegraphics[width=9cm]{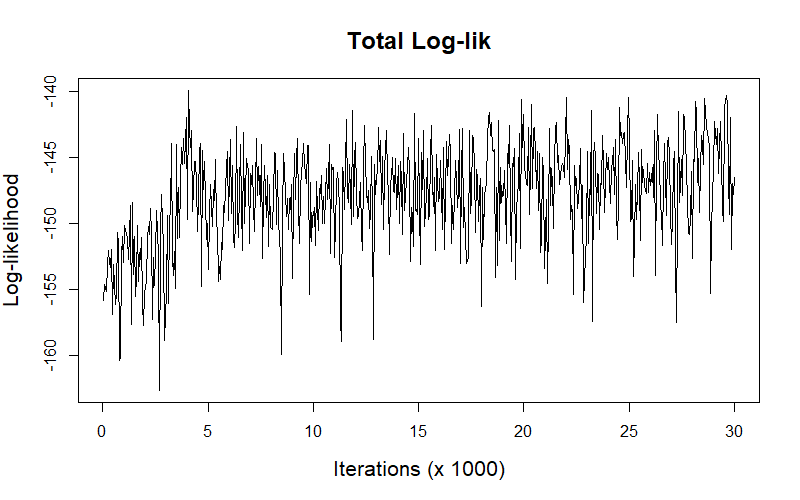}
    \caption{Complete log-likelihood as a function of iterations. The log-likelihood values are  depicted every 50 iterations.}
    \label{fig:LL_valid}
\end{figure}

\begin{figure}
    \centering
    \includegraphics[width=14cm]{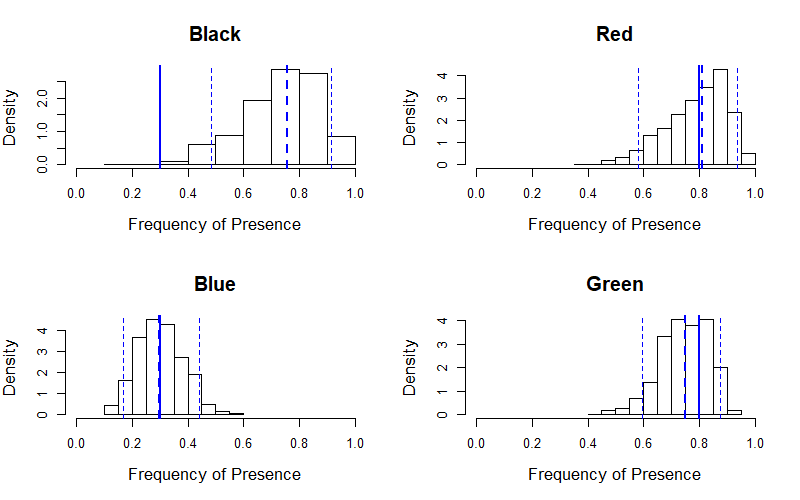}
    \caption{Posterior histograms of the frequencies $p_j$. Thick continuous line: true value of the parameter. Dashed thick line: posterior median. Dashed thin lines: posterior 0.05 and 0.95 quantiles.}
    \label{fig:Freq_Valid}
\end{figure}

\begin{figure}
    \centering
    \includegraphics[width=8cm]{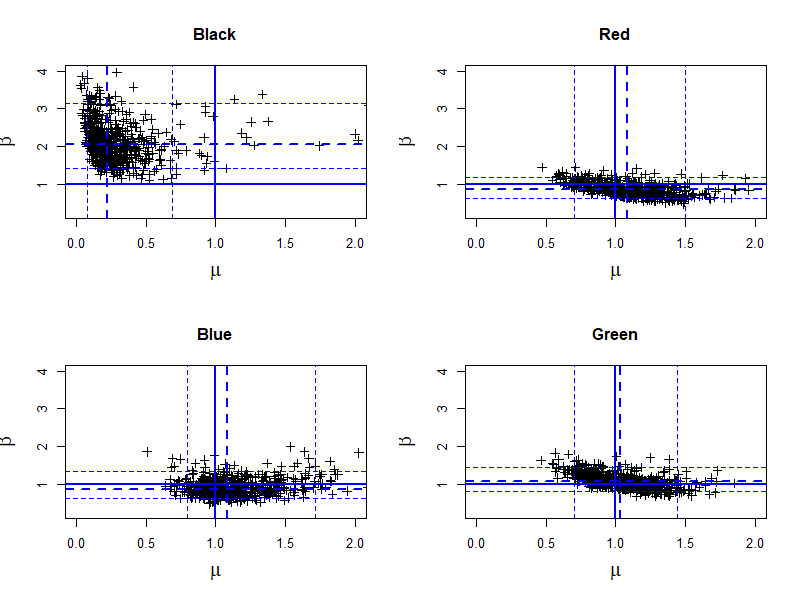} \
    \includegraphics[width=8cm]{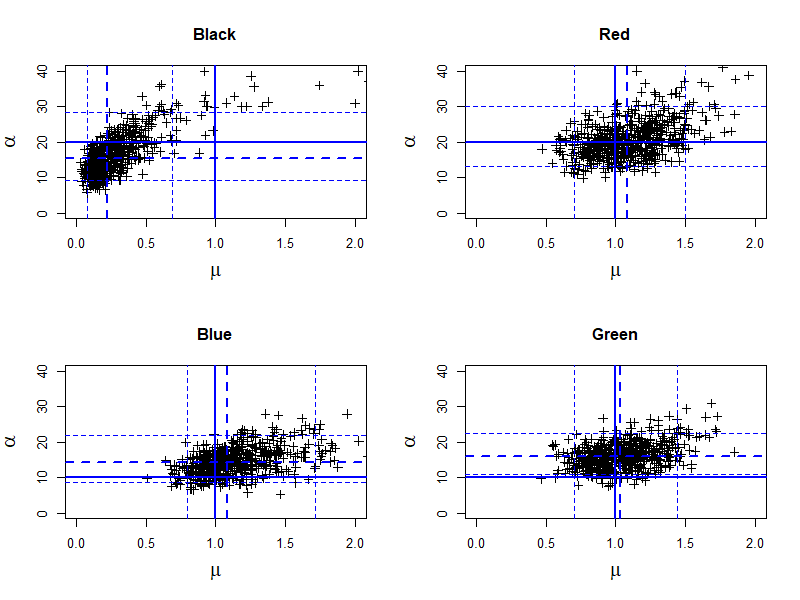}
    \caption{Left: posterior cross-plot of $\beta_j$ vs.  $\mu_j$. Right: posterior cross-plot of $\alpha_j$ vs.  $\mu_j$. Thick continuous line: true value of the parameter; Dashed thick line: posterior medians; dashed thin lines: posterior 0.05 and 0.95 quantiles.}
    \label{fig:Beta_vs_Mu_Valid}
\end{figure}

Figure \ref{fig:LL_valid} shows the complete log-likelihood as a function of the iterations. The mixing of the Markov chain is satisfactory and MCMC achieves convergence quite quickly. Figure \ref{fig:Freq_Valid}  shows the posterior distribution of the frequency of each category. At the exception of the \texttt{Black} category which was over-represented as already mentioned, the parameter $p_j$ is very well estimated. Figure \ref{fig:Beta_vs_Mu_Valid} shows the posterior cross-plot of the parameters $\beta_j$ (resp. $\alpha_j$) vs. $\mu_j$. One can see that there is some amount of negative correlation between $\beta_j$ and $\mu_j$, while there is some positive correlation between $\alpha_j$ and $\mu_j$. These findings are quite consistent with the parametric form of the function $\varphi(x) = \mu x^\beta$ on the one hand, and  with the result obtained in \cite{zhang2004inconsistent} regarding the simultaneous estimation of the range and variance of a Mat\'ern random field on the other hand. One can observe that the posterior median is quite close to the true value and always within the 90\% posterior credibility interval, at the exception of the range parameter for the \texttt{Black} category. For this category, it should be remembered that the observed frequency was over-represented (0.58, as compared to 0.3) and that the average thickness was 0.31 as compared to the theoretical expectation equal to 1.8. The maximum likelihood for the parameters ($p_j,\mu_j,\beta_j$) is thus completely off the real values ($0.3,1,1)$ as can also be seen on Fig. 6, where  $\mu_j$ is under-estimated and $\beta_j$ is over-estimated (Fig. 6). Nonetheless, given the good performances on the other categories, these results are quite promising considering that there are only 2 to 5 layers per category and that there are only 12 synthetic boreholes.

\subsection{Reconstruction of the sequences}

\begin{figure}
    \centering
    \includegraphics[width=16cm]{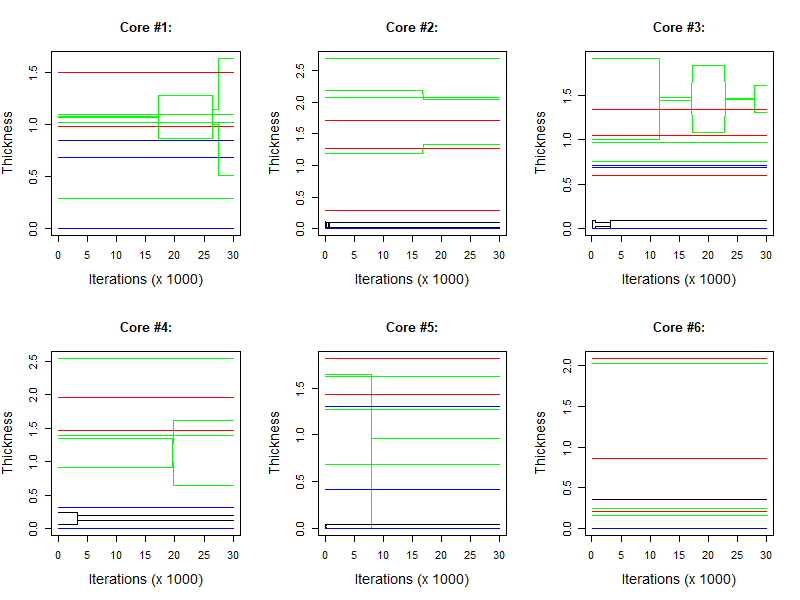}
    \caption{Thickness of different layers in synthetic boreholes $\sharp$1 to $\sharp$6 as a function of iterations. Layers are represented according to the color of the category they belong to.}
    \label{fig:cores}
\end{figure}

The observed sequence is not complete on most boreholes. Augmented sequences are created during the MCMC iterations. Since they can change along the iterations, the MCMC algorithm allow us to explore different consistent reconstructions. Figure \ref{fig:cores} shows the thickness of the 15 layers as a function of iterations for the first 6 synthetic boreholes. Each layer is color-coded according to its category. Similar plots were obtained for the other boreholes, but they are not shown here for the sake of concision.  Firstly, it should be noted that the thicknesses do not vary very often and that the variability of the thicknesses is quite different among the layers and among the boreholes. \texttt{Red} layers show constant thickness because in the parent sequence, \texttt{Red} layers are separated by 4, respectively 6 layers (see Fig. \ref{fig:sim}). As a consequence, the conditioning makes it impossible to \textit{Merge} or \textit{Split} any \texttt{Red} layers. The relative low number of moves is due to the quite strong lateral correlations implied by the smoothness parameter being equal to 3/2 and the range parameter being approximately equal to 1/3 of the size of the domain. On boreholes $\sharp$1 and $\sharp$6, there is no \texttt{Black} layer at all. The variations are not numerous and they concern mostly the 6-layer sequence [\texttt{Green}-\texttt{Blue}-\texttt{Green}-\texttt{Blue}-\texttt{Green}-\texttt{Blue}]   that allows some  exchanges of depth through successive moves. In particular, in boreholes $\sharp$1 and $\sharp$3 the actual sequence  is [\texttt{Green}-\texttt{Blue}-\texttt{Green}-\texttt{Blue}], so that some of the \texttt{Green} thickness can be exchanged between layers. Note that the total amount of \texttt{Green} thickness remains always constant. On boreholes $\sharp$2 to $\sharp$5, some \texttt{Black} layers are visible. The parent sequence is [\texttt{Black}-\texttt{Blue}-\texttt{Black}], but on {borehole} $\sharp$4 one of the observed thickness of \texttt{Blue} is 0. As a consequence, the observed \texttt{Black} thickness can be shared between the two layers, or it can be attributed to one layer only, the other one being zero. 

\begin{figure}
    \centering
    \includegraphics[width=16cm]{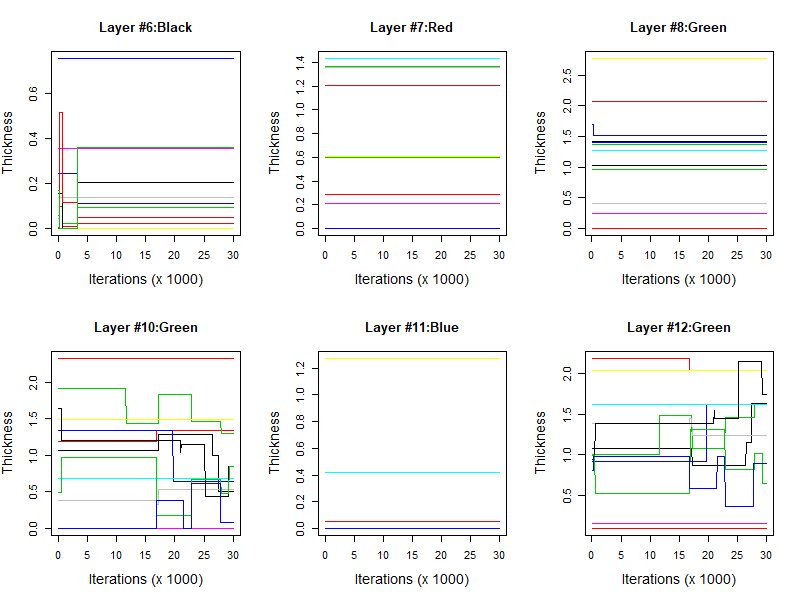}
    \caption{Thickness of layers $\sharp$6 to $\sharp$12 as a function of iterations. Each borehole is represented with a different color.}
    \label{fig:layers}
\end{figure}

Figure \ref{fig:layers} shows the thickness of layers $\sharp$6 to $\sharp$11 as a function of iterations for each borehole intersecting the layer. It is the dual representation of Fig. \ref{fig:cores}. Some layers have constant thickness across all boreholes, as it is the case for the \texttt{Red} layer $\sharp$7, which intersects 9 out of the 12 boreholes. On the three others, the conditioning does not make it possible to \textit{Merge} or \textit{Split} the layer. In layers $\sharp$10 and $\sharp$12 , the situation is quite the opposite. Since the total thickness must remain constant, variations on layers $\sharp$10 and $\sharp$12 are complementary for \texttt{Green}. These layers are part of the [\texttt{Green}-\texttt{Blue}-\texttt{Green}-\texttt{Blue}] sequence from layer 10 to layer 13 already mentioned.   This representation offers a complementary view of the variations of this layer.

\clearpage

\subsection{Conditional simulations}

\begin{figure}[hb]
    \centering
    \includegraphics[width=14cm]{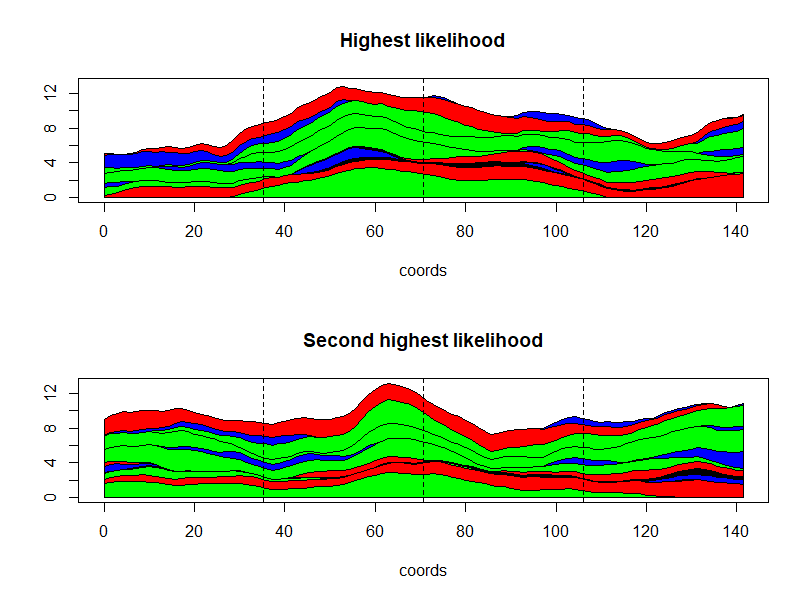}
    \caption{Two conditional simulations. The completed sequences and the posterior parameters correspond to the most likely configuration of the MCMC run.}
    \label{fig:CondSim}
\end{figure}

Two ingredients are necessary in order to perform a simulation conditional on the observed data. First, one needs all observed sequences to be coherently completed, in accordance with the parent sequence. Second, the simulation requires parameters for $\mu, \beta, p$ and $\alpha$. These must be jointly sampled from the posterior distribution in a coherent way. Independent and identically distributed sets of augmented sequences and estimated parameters are accessible by sampling from independently MCMC runs after the burn-in period. Alternatively, one can sample from the same MCMC run if the number of iterations between two samples is large enough. How much ``large enough'' is depends on the mixing properties of the MCMC algorithm. In practice, allowing a number of iterations larger than the burn-in period is a safe enough option. The set of parameters together with the completed sequences corresponding to the highest likelihoods recorded have been selected for conditional simulations. They are depicted in Fig. \ref{fig:CondSim}. Both simulations honor perfectly the data at the boreholes (dashed vertical lines), but they  show significantly different behaviors away from the conditioning data.

\section{A case study: deposition of materials on a aquifer}
\label{sec:Veneto}

\subsection{Study area and dataset description}
\begin{figure}
    \centering
    \includegraphics[width=14cm]{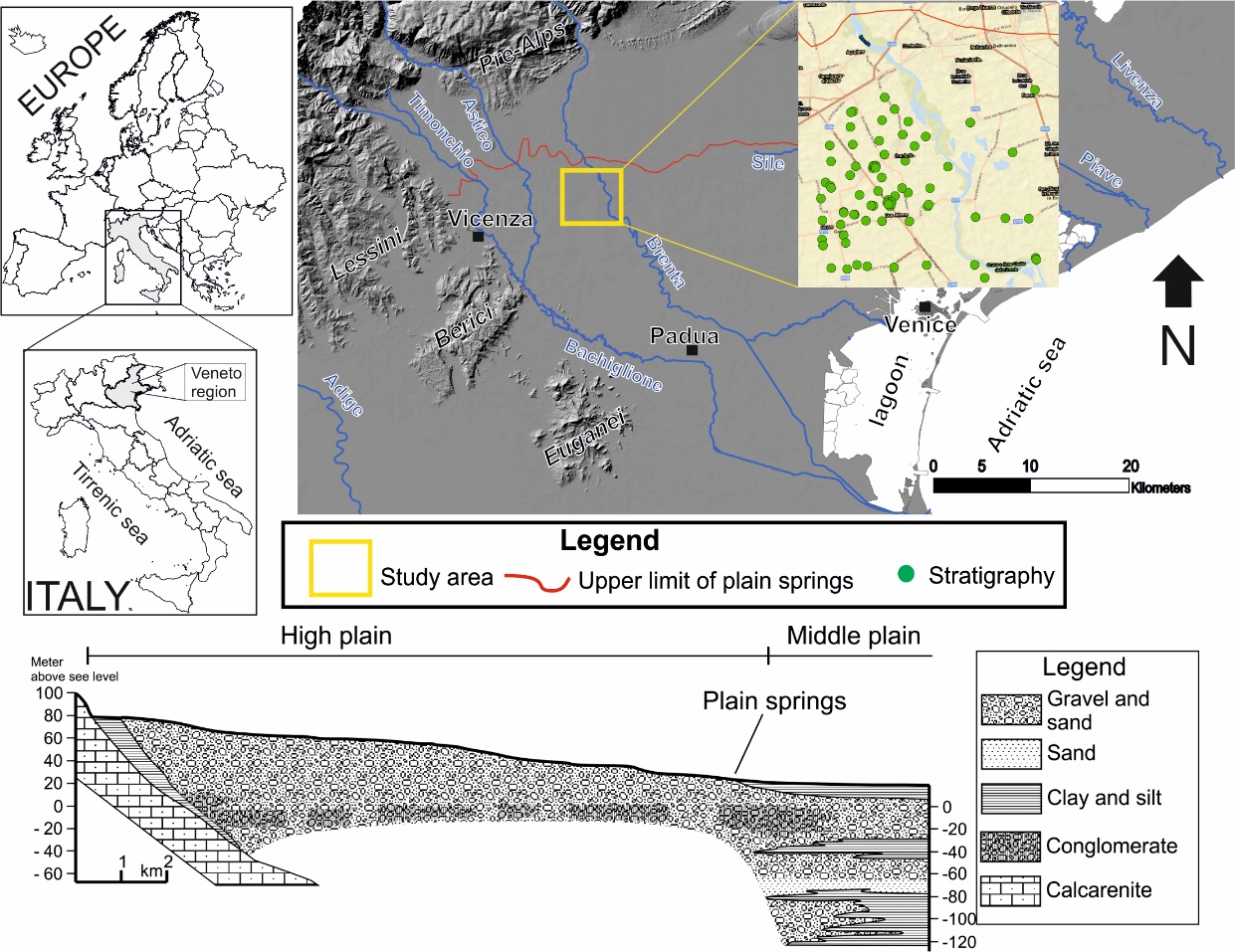}
    \caption{The study area of the real data example and the stratigraphy.}
    \label{fig:Veneto0}
\end{figure}

The study area  (Fig. \ref{fig:Veneto0}) is in the central part of the Venetian Plain (Italy), on the Brenta megafan (principally on the right bank of the actual Brenta River) of the Northern Padua district. In such an area several rivers (Bacchiglione, Brenta,  Astico and Timonchio) are responsible for the deposition of a significant portion of the material, hundreds of meters thick, which forms the subsoil of the Venetian Plain. Along the piedmont belt of the plain, fans from adjacent rivers laterally penetrate gravelly alluvial fans. The result is entirely gravelly subsoil throughout the thickness of the high Venetian Plain . Because deeper fans often invade further areas of the high plain from the undifferentiated gravel cover, the terminal parts of the fans extend downstream for various distances, producing an alluvial cover that is no longer uniformly gravely, but is instead composed by alternating layers of gravel and silty clay of swampy, lagoon or marine origin \citep{Fabbri-et-al2016}.

The data-set contains 24 {boreholes} drilled in a 5~km $\times$ 6~km region, with a minimum distance between {boreholes} of 0.23~km (Fig. \ref{fig:Veneto1}, top-left panel). Since the maximum depth of the {boreholes} is highly variable, a depth window between surface (from 35~m to 40~m above see level) and 25~m above see level is selected. There are four categories \texttt{L}(imo) (Silt), \texttt{S}(abbia) (Sand), \texttt{G}(hiaia) (Gravel), \texttt{A}(rgilla) (Clay) and the parent sequence, containing six layers, is: [\texttt{L}-\texttt{S}-\texttt{G}-\texttt{L}-\texttt{A}-\texttt{G}]. Notice that since there is only one layer for \texttt{S} and \texttt{A}, the associated thicknesses on the {boreholes} are known without ambiguity when present, which is not necessarily the case for the thicknesses associated to \texttt{L} and \texttt{G}. A range of 2 and 4 layers are observed on each {borehole}. One {borehole} contains an observed sequence of length 4 and 5 {boreholes} contain an observed sequence of length 3. The empirical estimates of the presence and the {average thicknesses} are shown in Table \ref{tab:empirical}. The most observed categories are \texttt{S} followed by \texttt{L}  as measured by the proportion of presence (for L, $\rho_j(0)=22/(2x24)=0.46$. The less observed category is \texttt{A}, with 3 records only. 

\begin{table}[htb]
    \centering
\begin{tabular}{l|ccccc}
        \hline\hline
             &   \texttt{L} & \texttt{S} & \texttt{G} & \texttt{A} & Overall \\
             \hline
Number of records & 22 & 18 & 12 &  3 & 55 \\
Proportion of presence, {$p_j(0)$}  &  0.46 &  0.75 & 0.25 &  0.13 & 0.38 \\
Average thickness  (in m), $\bar{T}_j$  &   0.73 &  2.25 & 3.89 &  1.10 &  1.94 \\
Initial value, $\tau_j(0)$ &  0.10 & -0.67 & 0.67 &  1.15& -- \\
initial value, $\mu_j(0)$ & 0.96 &  2.06 & 6.52 & 2.21 & -- \\
\hline\hline
\end{tabular}
    \caption{Empirical estimates of {presence and average thickness} and initial values for $\tau$ and $\mu$. \label{tab:empirical}}
\end{table}

\subsection{Model setting}
The empirical estimates are  transformed into initial values for $\tau_j$ and $\mu_j$, by setting initial values for $\beta_j$ to $\beta_j(0)=1$. Thus we get for each category $j$
$$\mu_j(0) = \frac{\bar{T}_j  {p_j(0)}}{\phi(\tau_j(0))},\quad  \hbox{with} \ \tau_j(0) = \Phi^{-1}(1- p_j(0)).$$
Preliminary tests (not reported here) showed that the likelihood computed with a Mat\'ern covariance function is almost always significantly larger  with a smoothness parameter $\nu = 1/2$ than with $\nu=3/2$  or $\nu = 5/2$. Therefore, the parameter $\nu$ is set to $1/2$, corresponding to an exponential covariance function, even though this covariance function corresponds to continuous but non differentiable random surfaces. This point will be further discussed in Sect. \ref{sec:conclusion}. Initial values for the range are set to  1~km.  

\begin{figure}
    \centering
    \includegraphics[width=5cm]{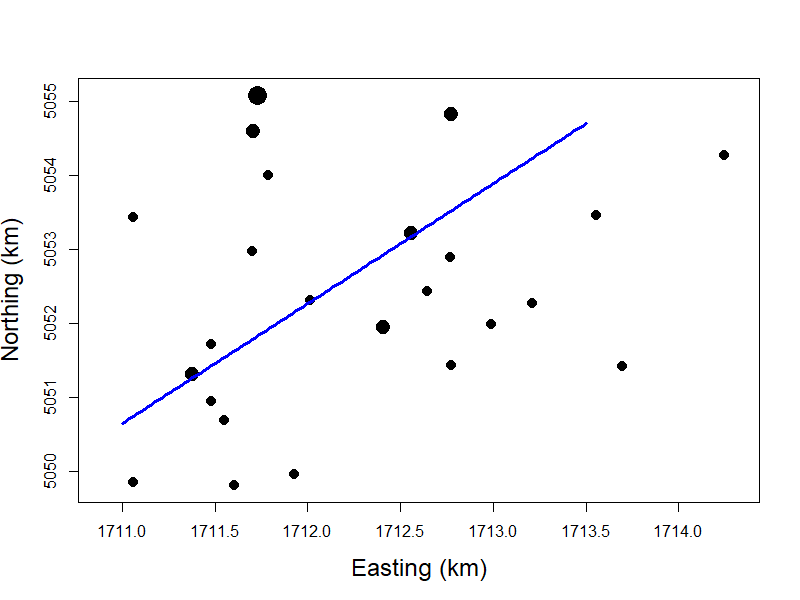}  \ \includegraphics[width=5cm]{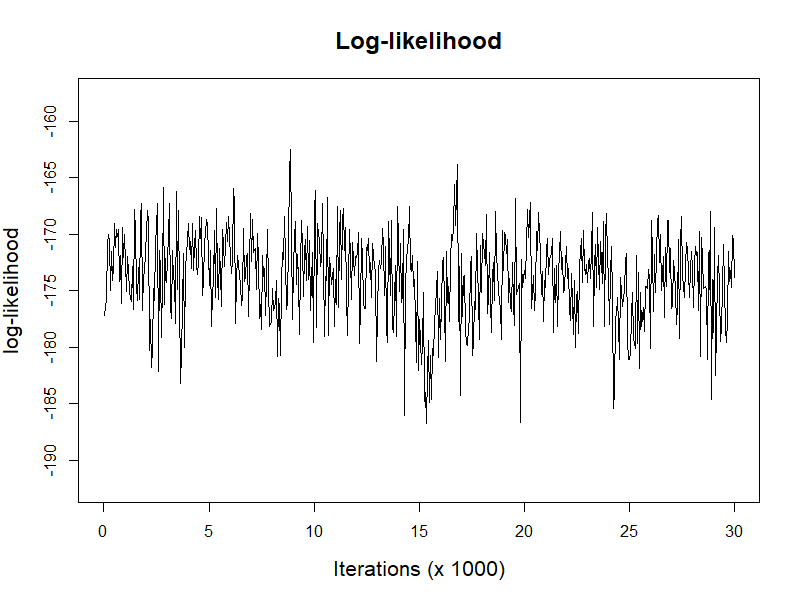} \ \includegraphics[width=5cm]{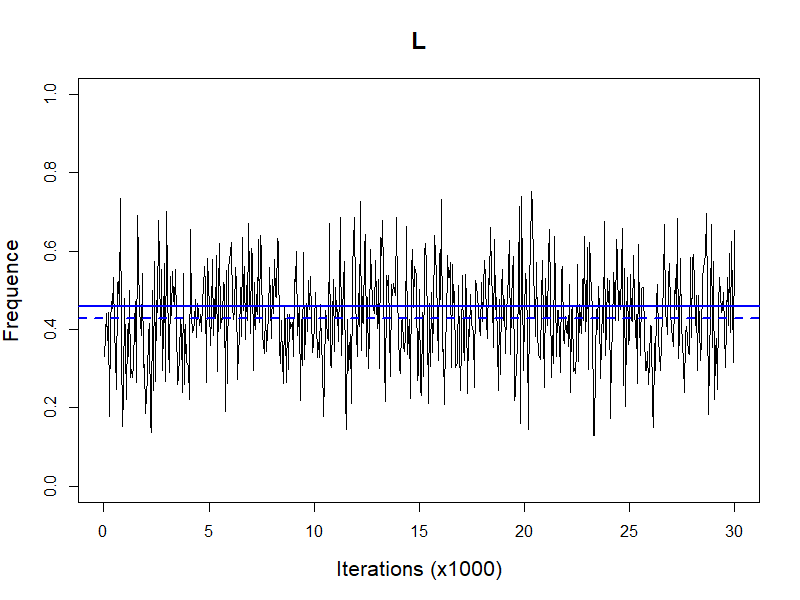} 
    
    \includegraphics[width=5cm]{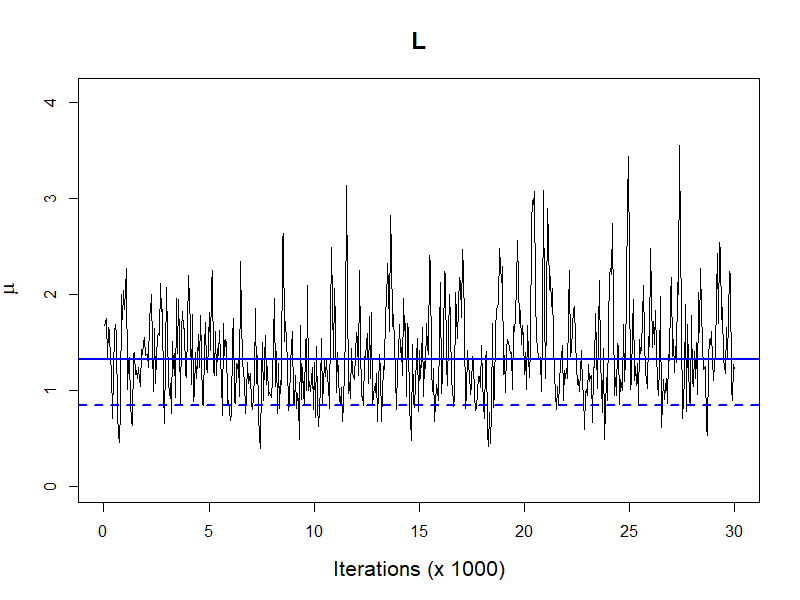} \ \includegraphics[width=5cm]{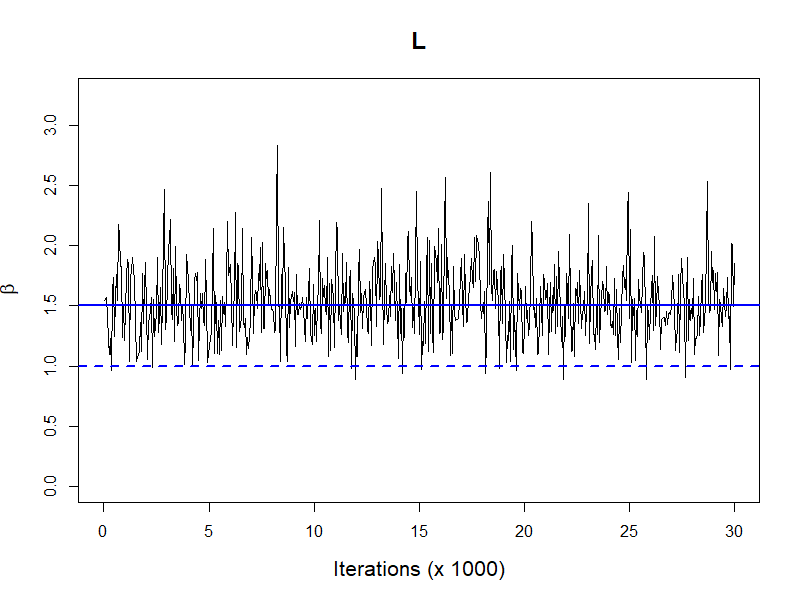} \ \includegraphics[width=5cm]{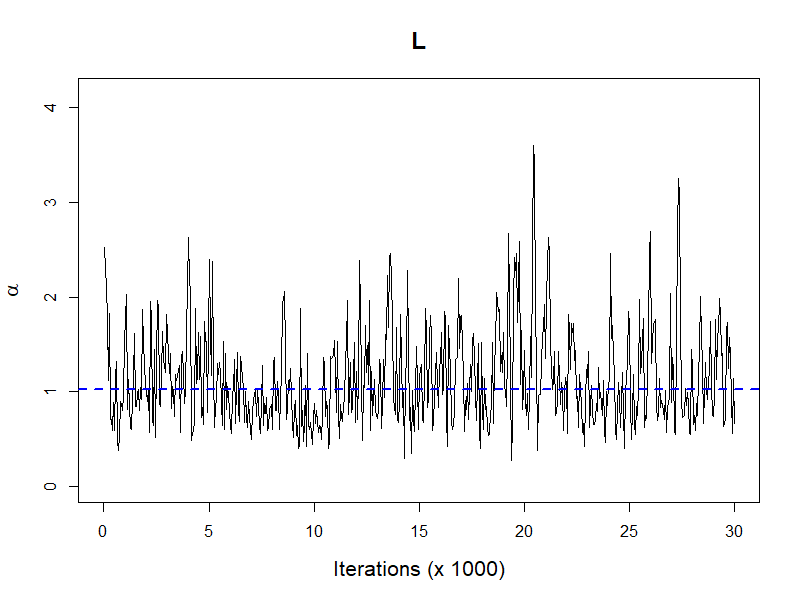} 
    \caption{Location of the 24 {boreholes} analyzed in the Veneto dataset (top left); diameter is proportional to the number of thicknesses recorded (from 2 to 4); thick blue line: cross-section for conditional simulation. Then, from top to bottom and from left to right: total likelihood, $p$, $\mu$, $\beta$ and {$\alpha$} as a function of iterations for category \texttt{L}. Continuous lines: posterior medians. Dashed lines: initial values.}
    \label{fig:Veneto1}
\end{figure}

In this dataset,  sequences are highly incomplete. As a consequence, the MCMC algorithm needs to have good mixing properties in order to explore the many possible augmented sequences that are compatible with the observations. Proposals follow a random walk with flat uninformative priors similar to that of Sect. \ref{sec:MCMC} for $p_j$ and $\beta_j$. PC priors were used for $\mu_j$ and $\alpha_j$, with $\epsilon_\alpha = \epsilon_\mu=0.01$ and $(\alpha_0,\mu_0) = (.25,10)$. Algorithm 2 is run for $30,000$ iterations, after a burn-in period of $2,500$  iterations. Values of parameters are then sampled every 50 iterations, so that $m=600$ posterior samples are collected. The proposals in the Metropolis-Hasting steps follow a uniform random walk with increments in $[-0.4,0.4]$ for $\mu_j$ and $\beta_j$, in $[-0.15,0.15]$ for $p_j$ and in $[-0.2,0.2]$  for the range $\alpha_j$. With these choices, the acceptance ratio for the parameters was around 0.8. 
Although it is higher than recommended, it does not appear to have a negative impact on the estimation procedure. Instead the acceptance ratio of new thickness configurations was equal to $0.22$ due to the incompleteness of this data set.
Figure \ref{fig:Veneto1} shows the values of the parameters $p$, $\mu$, $\beta$ and $\alpha$ as a function of iterations after burn-in, for category \texttt{L}. It is quite clear that the chain is stationary with good mixing. Notice the difference between the initial values and the posterior medians. Similar results have been obtained for the other categories.

\subsection{Results}

\label{sec:results}

\begin{figure}
    \centering
    \includegraphics[width=7.5cm]{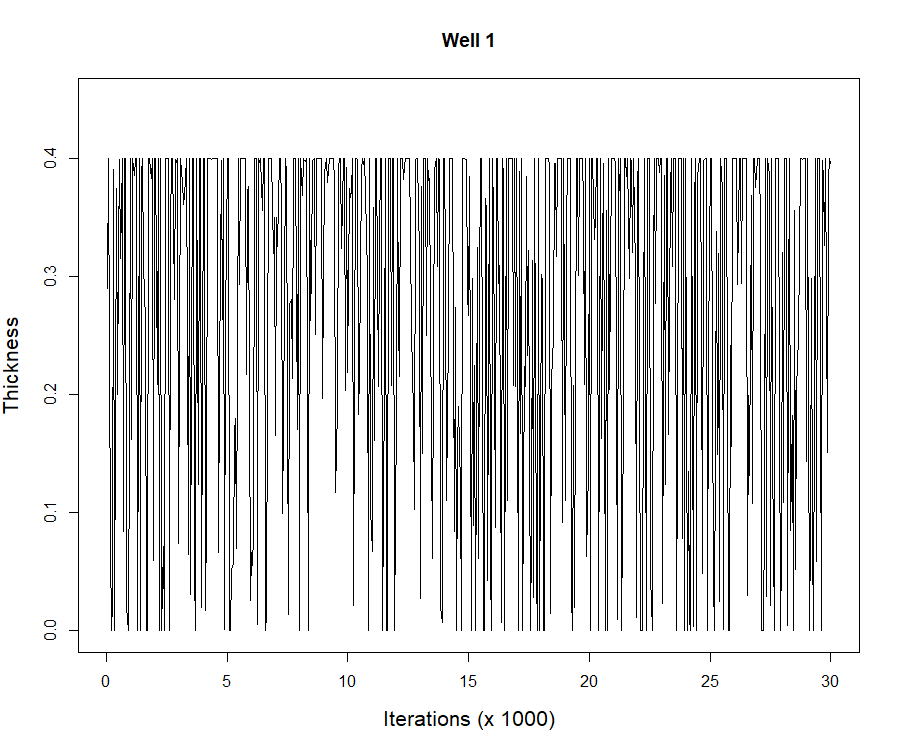} \qquad
    \includegraphics[width=7.5cm]{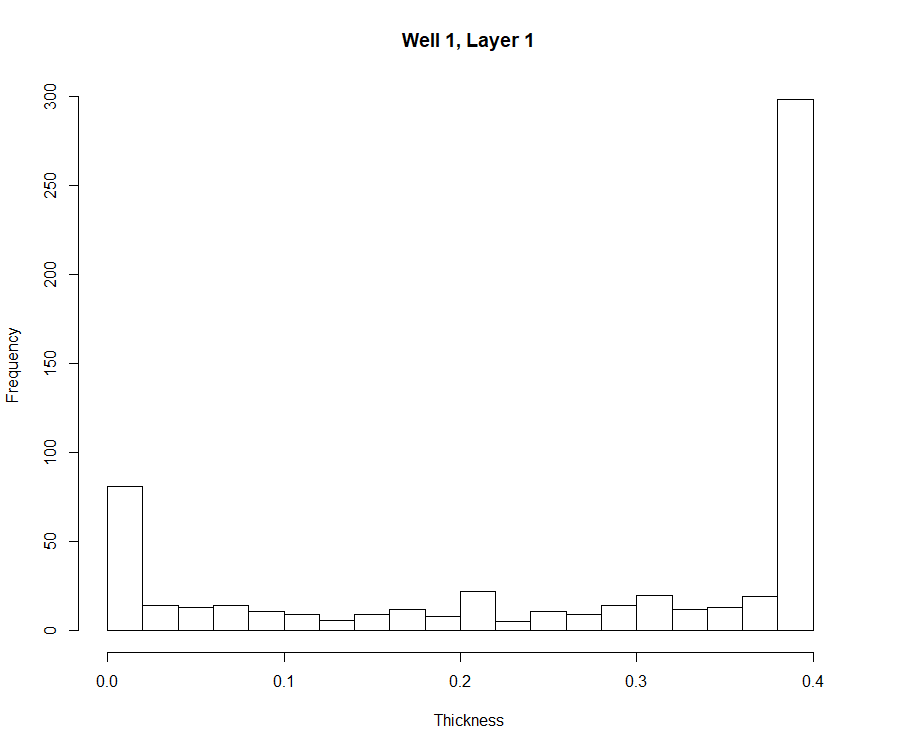}  
    \caption{Thickness of the first layer \texttt{L} in {borehole} \# 1. Left: as a function of iterations. Right: posterior histogram.}
\label{fig:Veneto2}
\end{figure}

\subsubsection*{Analysis of thicknesses}

When data belonging to the categories \texttt{L} and \texttt{G} are observed on the {boreholes}, the recorded thickness might belong to a single layer or to two layers. For these categories, the posterior thickness distribution might therefore look different to the observed one. Figure \ref{fig:Veneto2} (left) shows how thicknesses of the first layer \texttt{L} in {borehole} \#1 vary along iterations thanks to the \textit{ Split}, \textit{Merge} and \textit{Displace} moves of the MCMC. On this borehole, the observed sequence is [\texttt{L} - \texttt{A} - \texttt{G}]. The measured thickness for \texttt{L} is equal to $0.4$. Since the parent sequence is [\texttt{L} - \texttt{S} - \texttt{G} - \texttt{L} - \texttt{A} - \texttt{G}] this thickness could correspond to the first layer only (case I), to the fourth layer only (case II), or it could be shared between the two layers (case III). Figure \ref{fig:Veneto2} (right) represents the posterior histogram of the thickness in the first layer. Case I corresponds to $0.4$, case II to $0$ and case III to any value in the interval $(0,0.4)$. Frequencies computed along the iterations reveals that case III is the most likely case, with an estimated probability of $0.47$. The probabilities of case I and case II are equal to $0.42$ and $0.11$, respectively. Similar analysis can easily be performed on other {boreholes} and other categories.

\begin{figure}
    \centering
    \includegraphics[width=8cm]{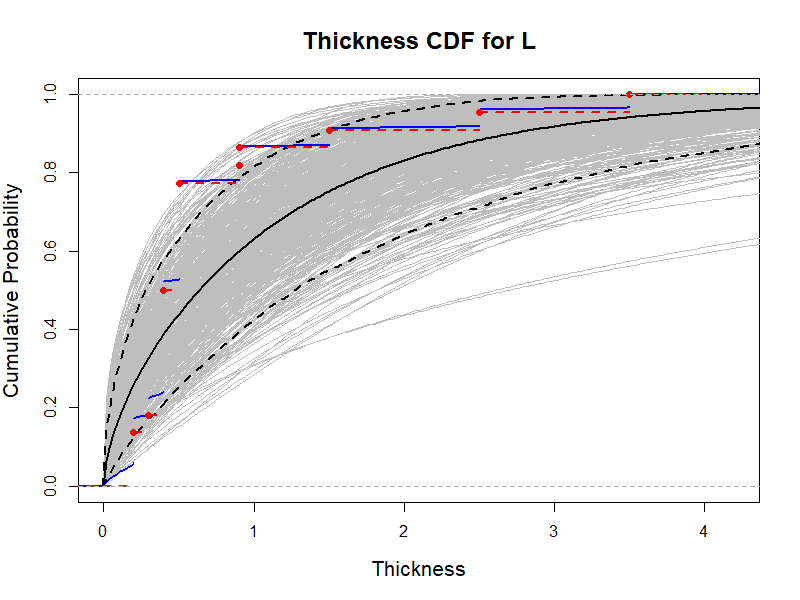} \
    \includegraphics[width=8cm]{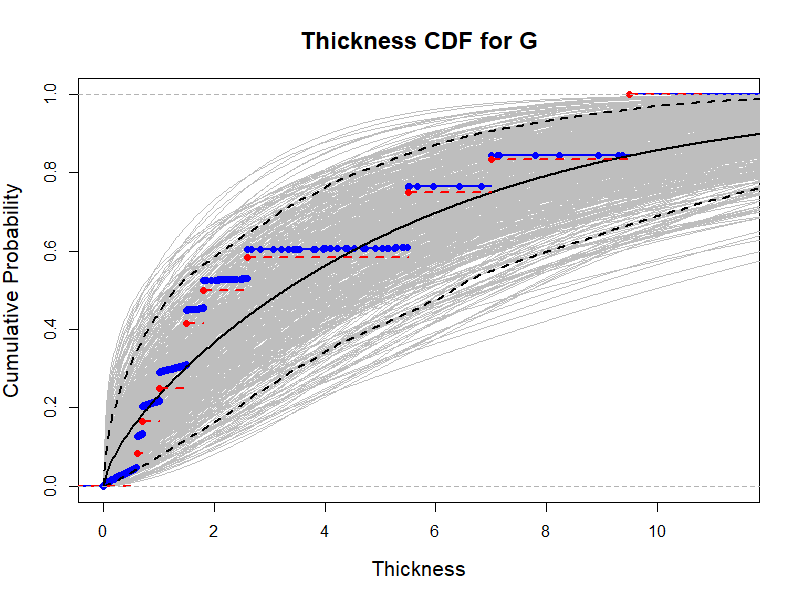}  
    \caption{Thickness Cumulative  Distributions (TCD). In gray: MCMC samples of the posterior theoretical TCD according to \eqref{eq:thick_cpf}; Black continuous curve: pointwise posterior median TCD; Black dashed curves: pointwise posterior $0.05$ and $0.95$ posterior quantiles. Red dashed curve: TCD of the original data; Blue curve: TCD of the MCMC samples.
    Left: category \texttt{L}; Right: category \texttt{G}.}
\label{fig:Veneto3}
\end{figure}

For a given category (for simplicity we drop the index $j$), and for given parameters $(p,\mu,\beta)$, the theoretical Thickness Cumulative Distribution (TCD) is:
\begin{equation}
    P(Z \leq z \mid p,\mu,\beta) =  \int_{\tau}^{\tau + (z/\mu)^{1/\beta}} \frac{\phi(y)}{p} dy = \frac{\Phi \left( \tau + (z/\mu)^{1/\beta} \right) - \Phi(\tau) }{p},
\label{eq:thick_cpf}
\end{equation}
with $\Phi(\tau)=1-p$. The parameters are sampled every 50 iterations of the MCMC, thereby mitigating the correlation between successive samples.  At each recorded iteration $k = 1,\dots,m$, the posterior samples $p(k)$, $\mu(k)$ and $\beta(k)$ make it possible to compute a posterior theoretical TCD according to  \eqref{eq:thick_cpf}. Those are represented in gray on Fig. \ref{fig:Veneto3}  for categories \texttt{L} and \texttt{G}. The ensemble of $m$ posterior TCDs allows us to compute pointwise median and the pointwise quantiles $q_{0.05}$ and $q_{0.95}$, which  are represented with black continuous and dashed lines, respectively.

Empirical posterior TCD can alternatively be computed from the thickness values recorded along the sampled iterations $k=1,\dots,m$. In principle, empirical and theoretical TCDs should match. Figure \ref{fig:Veneto3} shows  the original and posterior TCDs, respectively in red and blue.  Thanks to the \textit{Split}, \textit{Merge} and \textit{Displace} movements, the posterior TCD is slightly smoother than the original one since  values intermediate to the observed ones are simulated.

Overall, the match between the empirical and the theoretical TCD is very satisfactory since the empirical curve is fully included in the envelope of the MCMC samples for category \texttt{G} and mostly included in the envelope for category \texttt{L}.

\subsubsection*{Spatial analysis and conditional simulation}

\begin{figure}
    \centering
    \includegraphics[width=14cm]{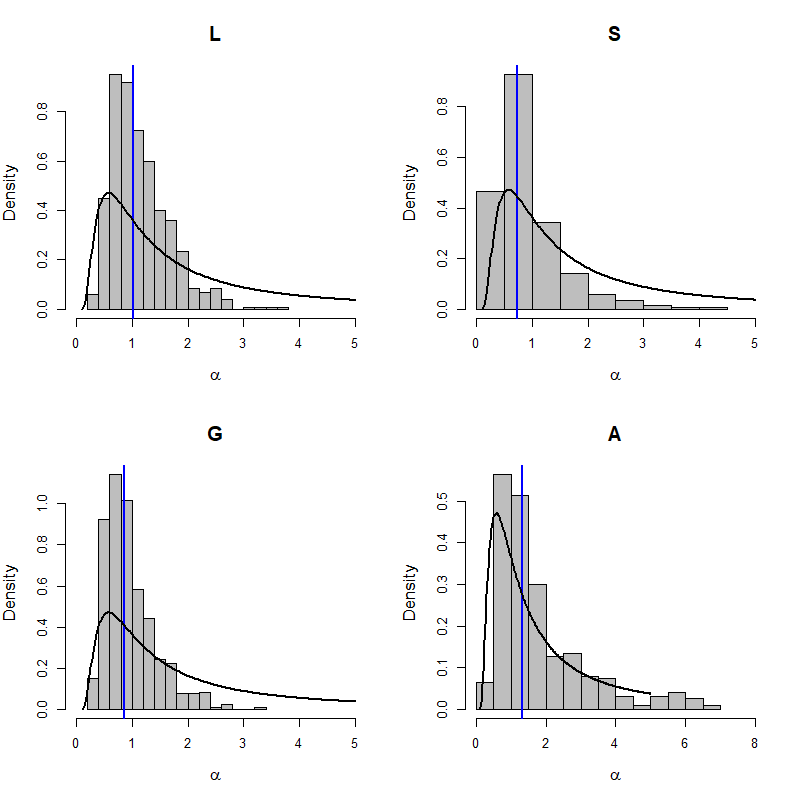}   
    \caption{For each category, posterior histogram of the spatial range and prior distribution (continuous line). Blue vertical line: posterior median.}
\label{fig:Veneto4}
\end{figure}

Figure \ref{fig:Veneto4} shows the posterior histograms of the spatial range for the four categories, with the prior density being also shown. This figure indicates that the prior has a heavy weight on the posterior distributions for each unit. However, when a category is well informed (\texttt{L} and \texttt{G}), the posterior distribution is more concentrated around the posterior median (indicated with a vertical blue line), equal to $1.03$, $0.73$ and $0.85$ for categories \texttt{L}, \texttt{S} and \texttt{G}, respectively. On the contrary, category \texttt{A} has only three records. Since there is very little information in the likelihood, the posterior distribution is very close to the prior one. The result of this analysis is that there is indeed a significant amount of spatial correlations in the random fields modeling the thickness of the layers for all categories but \texttt{A}. 

\begin{figure}
    \centering
    \includegraphics[width=14cm]{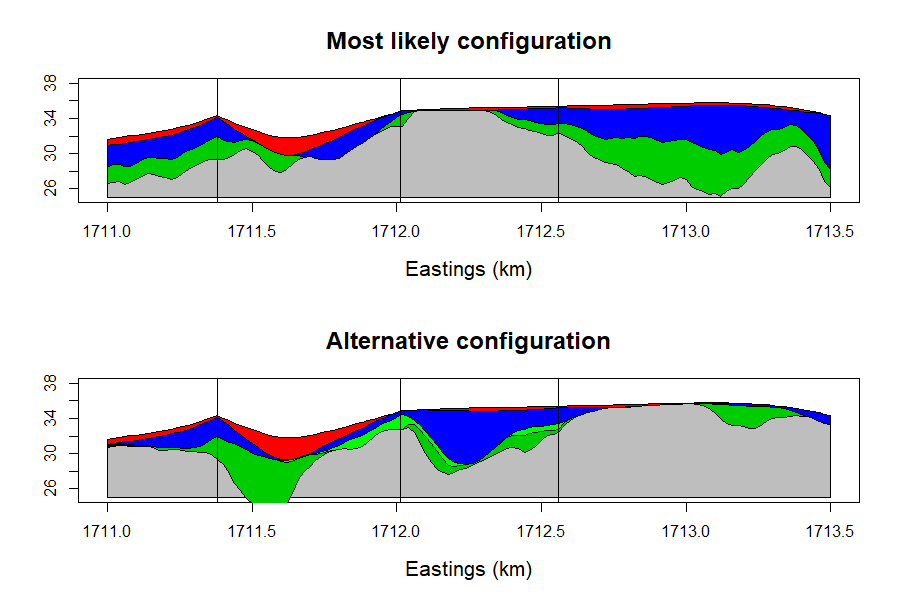}   
    \caption{Two cross-sections along the line shown in Fig. \ref{fig:Veneto1} (top left). Notice that there are two different layers for \texttt{G} in the bottom cross-section.}
    
\label{fig:Veneto5}
\end{figure}

Figure \ref{fig:Veneto5} shows two conditional simulations performed along the cross-section depicted in Fig. \ref{fig:Veneto1} (top left). This cross-section has been chosen because it is close to three conditioning {boreholes} (shown with black vertical lines on Fig. \ref{fig:Veneto5}) with incomplete observed sequences that allow different thickness configurations in category \texttt{G}. The color code is the following: red for \texttt{L}, blue for \texttt{S}, green for \texttt{G} and black for \texttt{A}. The gray color corresponds to undefined lithofacies below the last recorded layer. The first cross-section corresponds to iteration $8,900$ after burn-in, for which the likelihood was the highest along the whole MCMC (log-likelihood is equal to $-162.5$). Here, the \texttt{G} thickness is entirely in layer \# 6. The second cross-section corresponds to a configuration where the \texttt{G} thickness is now shared between the two layers. Different shades of green have been used to distinguish the two layers. This second configuration corresponds to the most likely configuration with shared thicknesses between the two  \texttt{G} layers (log-likelihood is equal to $-171.3$). Notice that it is significantly less likely than the first configuration, indicating that the data is orders of magnitude less likely with the second configuration than with the first one. Notice also that the cross-sections are quite different away from the conditioning {boreholes}. The parameters corresponding to these two configurations are reported in Table \ref{tab:2config}.

\begin{table}[htb]
    \centering
\begin{tabular}{lrrrrrrrrr}
\hline\hline
&  \multicolumn{4}{c}{First configuration} & \phantom{XX} & \multicolumn{4}{c}{Second configuration} \\
&  \multicolumn{4}{c}{Log-likelihood $= -162.5$} & \phantom{XX} & \multicolumn{4}{c}{Log- likelihood $= -171.3$} \\
\hline
&   \multicolumn{1}{c}{\texttt{L}} & \multicolumn{1}{c}{\texttt{S}} & \multicolumn{1}{c}{\texttt{G}} & \multicolumn{1}{c}{\texttt{A}} & & \multicolumn{1}{c}{\texttt{L}} & \multicolumn{1}{c}{\texttt{S}} & \multicolumn{1}{c}{\texttt{G}} & \multicolumn{1}{c}{\texttt{A}} \\
\hline
$p_j$  &  0.40 &  0.81 & 0.23 &  0.05 & & 0.45 & 0.64 & 0.46 & 0.48 \\ 
$\mu_j$ & 1.29 & 1.80 & 6.99 & 2.20 & & 1.98 & 2.02 & 5.01 & 11.03 \\
$\beta_j$ & 1.54 & 1.59 & 1.01 & 0.50 & & 1.42 & 1.41 & 1.39 & 1.76 \\
$\alpha_j$ & 1.25 & 0.29 & 0.78 & 0.71 & & 2.03 & 0.54 & 0.43 & 3.58 \\
\hline\hline
\end{tabular}
    \caption{Parameters corresponding to the two configurations \label{tab:2config} shown in Fig. \ref{fig:Veneto5}.}
\end{table}

\section{Concluding remarks}
\label{sec:conclusion}
In this paper  a new rule-based approach for simulating depositional sequences of surfaces conditionally to lithofacies thickness data has been presented. A distinctive feature of this approach is that it takes properly into account  the different amount of information along horizontal and vertical dimensions usually contained in  borehole datasets: few cores and, consequently, few horizontal information but complete information along the depth.

This is achieved by supposing that there exists a common lithological sequence of facies, compatible with the observed data. Moreover the sequence is supposed to be known in advance. The facies thickness, which is  non-negative, is modeled by means of a truncated and transformed stationary Gaussian field. In principle other non-negative random fields could be considered, but this choice allowed us to exploit the flexibility of Gaussian random fields in the selection of the covariance functions with different degree of smoothness. Evaluation of the likelihood is made possible thanks to the Gaussian framework for which well known methods and efficient computing tools are available. 
 
A  data augmentation  algorithm, coupled with a MCMC algorithm, is employed for learning the parameters of the stochastic model from borehole data. A very interesting feature of the proposed algorithm is that it allows to explore all different configurations that are compatible with the available data. Thanks to the MCMC approach and the Bayesian framework it associates a likelihood to each of the possible realization corresponding to a set of parameters. From those, as shown in Sect. \ref{sec:results}, one can assess an empirical probability for each different configuration, select the most likely configurations and compute many other statistics of interest to the user.  The algorithm  requires multiple (to the order of $M \times n$) evaluation of the joint probability of a Gaussian vector being below a given threshold.  The current implementation in R uses the  \texttt{mvtnorm} package \citep{GenzPackage} that handles rather easily vectors with a few dozens of coordinates. It starts to slow down quite significantly around 100 coordinates and is unable to cope with more than 1000 coordinates. Further research is thus required if the number of boreholes goes from moderate to high or very high. One possible choice could be the approximation proposed in \cite{martinetti2017approximate}, but the impact of using a less precise approximation remains to be evaluated.

A too small dataset entails difficulties in specifying the regularity and the range of the covariance function, as was shown with category \texttt{A} that has only three records. It was found in the present work that parameters were reasonably well estimated with 15 records per category. On the other hand, as the data set gets larger and denser (for example when the horizontal distance between nearest neighbor boreholes becomes a small fraction of the range parameter) the likelihood will get more peaked around local maxima, thereby decreasing the mixing of the MCMC. In this case exploring all configurations coherent with the parent sequence is likely to become more difficult. Longer chains and multiple chains starting from very different initial configurations will probably be necessary.

Several assumptions and restrictions have been made in this work, which can be lifted in order to generalize this work. The stationarity assumption, which in the example considered here has proved appropriate, could be relaxed and the parameters could be easily modified to take covariates into account. Only a few half-integer values of the smoothness parameters have been considered, and the fitting of this parameter was done outside the MCMC machinery.  In principle the smoothness parameter could be different for different facies and it could be estimated in the Bayesian framework, just as any other parameter. Estimating simultaneously the three parameters of the Mat\'ern covariance in a Bayesian context is known to be extremely difficult. When there are only few data, this was made possible thanks to the PC priors \citep{fuglstad2019constructing}. Currently, to the best of our knowledge, the simultaneous PC prior for $(\nu, \alpha, \sigma^2)$ for Mat\'ern covariance is not known. We leave it for further research to find such PC priors.

Currently,  independent MCMCs are launched, one for every possible value $\nu\in\{1/2,3/2,5/2\}$. The one with the highest likelihood and the best mixing is selected and $\nu$ is fixed at that value. When analyzing the data from the Venetian plain, it was found that $\nu=1/2$ was best, despite the fact that the associated thicknesses (and thus surfaces) are mean-square continuous but not differentiable. One could have impose $\nu=3/2$, but at the cost of a very short spatial range implying almost no spatial correlation.  Whether one should let the data speak or impose a model for the regularity is a  debate. Here, the choice was to be guided by the data.

{Finally, the function that transform the Gaussian values to thicknesses was chosen to be a power function, but any other positive function could be used.

One information that is often available in real applications and on much more points than boreholes is the nature of the facies on surface. It is possible to incorporate such information at the cost of small changes in the method. At a given location $s$ where this information is available, one could consider that the facies of the upper layer, say facies $j$, is known and has positive thickness. The conditioning data would therefore be that $W_{\rm upper}(x)>\tau_j$. This conditioning can easily be handled within our MCMC procedure. At this location, there would be no conditioning for the other layers. 

The proposed approach depends on the existence and on the knowledge of a common lithological sequence of facies compatible with the observed data. If the sequence is unknown,  it is possible to derive it from the data, possibly by imposing some restriction such as minimum length for example. This problem has not been tackled here, since it has been evaluated out of the scope of this work. However it is worth to mention that the approach presented here can be modified to take into account several different parent sequences with their associated prior probabilities.

\subsection*{Acknowledgment}
This work was initiated during a visit of the first author to Ca' Foscari University of Venice . He acknowledges the support of  that institution. We wish to thank two anonymous reviewers for their in-depth and detailed reading of the first version of the manuscript. Their  many valuable comments  helped us to improve the manuscript.

\clearpage
\appendix
\section{A longer example of incomplete sequence}

\begin{table}[ht]
\caption{A longer and more complex example of a parent sequence $\bC$=[\texttt{Blue}-\texttt{Red},
	\texttt{Blue}-\texttt{Green}-\texttt{Blue}-\texttt{Red}-\texttt{Green}-\texttt{Blue}] with respect to  a recorded sequence $\bC^o$ and $\bT^o$. Only nine compatible augmented sequences are reported. \label{tab:ex2}}
	
	{\small
	\begin{center}
		
		\begin{tabular}{ccc|ccc|ccc|ccc}
			\hline\hline
			Parent & \multicolumn{2}{c|}{Recorded} & \multicolumn{9}{c}{Compatible augmented sequences}\\
			\hline
			$\bC$ & $\bC^o$ & $\bT^o$ &$\bC^a$& $\bT^a$  & $\bZ^a$ &$\bC^a$ & $\bT^a$  & $\bZ^a$ &$\bC^a$ &
			$\bT^a$  & $\bZ^a$ \\
			\hline\hline 
			\texttt{Blue} & \texttt{Blue}  & $T^o_1$ & \texttt{Blue} & $T^o_1$ & $T^o_1$     & \texttt{Blue} & $T^o_1$ & $T^o_1$     & \texttt{Blue} & $T^o_1$ & $T^o_1$ \\ 
			\texttt{Red} & \texttt{Red}  & $T^o_2$ & \texttt{Red} & $T^o_2$ & $T^o_2-T^o_1$ & \texttt{Red} & $T^o_2$ & $T^o_2-T^o_1$ &   & $T^o_1$ & 0\\
			\texttt{Blue} & \texttt{Green}   & $T^o_3$ &  & $T^o_2$  & 0         &   & $T^o_2$ & 0         &   & $T^o_1$  & 0\\
			\texttt{Green}  & \texttt{Blue}  & $T^o_4$ & \texttt{Green}  & $T^o_3$ & $T^o_3-T^o_2$ &   & $T^o_2$ & 0         &   & $T^o_1$  & 0 \\
			\texttt{Blue} & -- & --    & \texttt{Blue} & $T^o_4$ & $T^o_4-T^o_3$ &   & $T^o_2$ & 0         &   & $T^o_1$  & 0\\
			\texttt{Red} & -- & --    &   & $T^o_4$ &  0        &   & $T^o_2$ & 0         & \texttt{Red} &  $T^o_2$ & $T^o_2-T^o_1$\\
			\texttt{Green}  & -- & --    &   & $T^o_4$ &  0        & \texttt{Green}  & $T^o_3$ & $T^o_3-T^o_2$ & \texttt{Green}  & $T^o_3$  & $T^o_3-T^o_2$ \\
			\texttt{Blue} & -- & --    &   & $T^o_4$ &  0        & \texttt{Blue} & $T^o_4$ & $T^o_4-T^o_3$ & \texttt{Blue} & $T^o_4$  & $T^o_4-T^o_3$ \\
			\hline\hline 
			 &  &  & \texttt{Blue}  & $T^o_1$ & $T^o_1$     & \texttt{Blue} & $T^o_1$ & $T^o_1$     & \texttt{Blue} & $T^o_1$ & $T^o_1$ \\ 
			 &  &  & \texttt{Red}  & $\tilde{T}$ & $\tilde{T}-T^o_1$ & \texttt{Red} & $T^o_2$ & $T^o_2-T^o_1$ &   & $T^o_1$ & 0\\
			 &  &  &    & $\tilde{T}$ & 0         &   & $T^o_2$ & 0          &  \texttt{Red} & $T^o_2$  &  $T^o_2-T^o_1$\\
			 &  &  &    & $\tilde{T}$ & 0         & \texttt{Green}  & $\tilde{T}$ & $\tilde{T}-T^o_2$  &  \texttt{Green}  & $T^o_3$  &  $T^o_3-T^o_2$ \\
			 &  &  &    & $\tilde{T}$ & 0         &   & $\tilde{T}$ & 0          &  \texttt{Blue} & $\tilde{T}$  &  $\tilde{T}-T^o_3$\\
			 &  &  &  \texttt{Red} & $T^o_2$ & $T^o_2-\tilde{T}$ &   & $\tilde{T}$ & 0          &  &  $\tilde{T}$ & 0\\
			 &  &  &  \texttt{Green}  & $T^o_3$ & $T^o_3-T^o_2$ & \texttt{Green}  & $T^o_3$ & $T^o_3-\tilde{T}$  &  & $\tilde{T}$  & 0 \\
			 &  &  &  \texttt{Blue} & $T^o_4$ & $T^o_4-T^o_3$ & \texttt{Blue} & $T^o_4$ & $T^o_4-T^o_3$  & \texttt{Blue} & $T^o_4$  & $T^o_4-\tilde{T}$ \\
			\hline\hline 
			 &  &  &  \texttt{Blue} & $\tilde{T}$ & $\tilde{T}$      & \texttt{Blue} & $\tilde{T}$ & $\tilde{T}$     & \texttt{Blue} & $\tilde{T}$ & $\tilde{T}$ \\ 
			 &  &  &    & $\tilde{T}$ & 0          &   & $\tilde{T}$ &  0         &   & $\tilde{T}$ & 0\\
			 &  &  &    & $\tilde{T}$ & 0          & \texttt{Blue}  & $T^o_1$ & $T^o_1-\tilde{T}$ &  \texttt{Blue} & $\tilde{\tilde{T}}$  & $\tilde{\tilde{T}}-\tilde{T}$ \\
			 &  &  &    & $\tilde{T}$ & 0          &    & $T^o_1$ & 0          &   & $\tilde{\tilde{T}}$  & 0 \\
			 &  &  &  \texttt{Blue} & $T^o_1$ & $T^o_1-\tilde{T}$  &    & $T^o_1$ & 0          &  \texttt{Blue} & $T^o_1$  & $T^o_1-\tilde{\tilde{T}}$\\
			 &  &  &  \texttt{Red} & $T^o_2$ & $T^o_2-T^o_1$  & \texttt{Red}  & $T^o_2$ & $T^o_2-T^o_1$  &  \texttt{Red} &  $T^o_2$ & $T^o_2-T^o_1$\\
			 &  &  &  \texttt{Green}  & $T^o_3$ & $T^o_3-T^o_2$  & \texttt{Green}   & $T^o_3$ & $T^o_3-T^o_2$  &  \texttt{Green}  & $T^o_3$  & $T^o_3-T^o_2$ \\
			 &  &  &  \texttt{Blue} & $T^o_4$ &  $T^o_4-T^o_3$ & \texttt{Blue}  & $T^o_4$ & $T^o_4-T^o_3$  & \texttt{Blue} & $T^o_4$  & $T^o_4-T^o_3$ \\
			\hline\hline 
		\end{tabular}
	\end{center}
	}
\end{table}

\clearpage

\bibliographystyle{apalike}

\begin{thebibliography}{}
	
	\bibitem[Allard et~al., 2011]{allard2011efficient}
	Allard, D., D'Or, D., and Froidevaux, R. (2011).
	\newblock An efficient maximum entropy approach for categorical variable
	prediction.
	\newblock {\em European Journal of Soil Science}, 62(3):381--393.
	
	\bibitem[Allard et~al., 2006]{allard2006conditional}
	Allard, D., Froidevaux, R., and Biver, P. (2006).
	\newblock Conditional simulation of multi-type non stationary Markov object
	models respecting specified proportions.
	\newblock {\em Mathematical Geology}, 38(8):959--986.
	
	\bibitem[Allcroft and Glasbey, 2003]{allcroft2003latent}
	Allcroft, D.~J. and Glasbey, C.~A. (2003).
	\newblock A latent Gaussian Markov random-field model for spatiotemporal
	rainfall disaggregation.
	\newblock {\em Journal of the Royal Statistical Society: Series C (Applied
		Statistics)}, 52(4):487--498.
	
	\bibitem[Armstrong et~al., 2011]{armstrong2011plurigaussian}
	Armstrong, M., Galli, A., Beucher, H., Loc'h, G., Renard, D., Doligez, B.,
	Eschard, R., and Geffroy, F. (2011).
	\newblock {\em Plurigaussian Simulations in Geosciences}.
	\newblock Springer.
	
	\bibitem[Baxevani and Lennartsson, 2015]{baxevani2015spatiotemporal}
	Baxevani, A. and Lennartsson, J. (2015).
	\newblock A spatiotemporal precipitation generator based on a censored latent
	Gaussian field.
	\newblock {\em Water Resources Research}, 51(6):4338--4358.
	
	\bibitem[Benoit et~al., 2018a]{benoit2018stochastic}
	Benoit, L., Allard, D., and Mariethoz, G. (2018a).
	\newblock Stochastic rainfall modeling at sub-kilometer scale.
	\newblock {\em Water Resources Research}, 54(6):4108--4130.
	
	\bibitem[Benoit et~al., 2018b]{benoit2018directional}
	Benoit, N., Marcotte, D., Boucher, A., D?Or, D., Bajc, A., and Rezaee, H.
	(2018b).
	\newblock Directional hydrostratigraphic units simulation using MCP algorithm.
	\newblock {\em Stochastic Environmental Research and Risk Assessment},
	32(5):1435--1455.
	
	\bibitem[Bertoncello et~al., 2013]{bertoncello2013conditioning}
	Bertoncello, A., Sun, T., Li, H., Mariethoz, G., and Caers, J. (2013).
	\newblock Conditioning surface-based geological models to well and thickness
	data.
	\newblock {\em Mathematical Geosciences}, 45(7):873--893.
	
	\bibitem[Beucher et~al., 1993]{beucher1993including}
	Beucher, H., Galli, A., Le~Loc?h, G., Ravenne, C., Group, H., et~al. (1993).
	\newblock Including a regional trend in reservoir modelling using the truncated
	Gaussian method.
	\newblock In Soares, editor, {\em Geostat Tr{\'o}ia '92}, pages 555--566.
	Kluwer.
	
	\bibitem[Carle and Fogg, 1996]{carle1996transition}
	Carle, S.~F. and Fogg, G.~E. (1996).
	\newblock Transition probability-based indicator geostatistics.
	\newblock {\em Mathematical Geology}, 28(4):453--476.
	
	\bibitem[Chil\`es and Delfiner, 2012]{Chiles2012}
	Chil\`es, J.-P. and Delfiner, P. (2012).
	\newblock {\em Geostatistics: Modeling Spatial Uncertainty}.
	\newblock John Wiley \& Sons, 2nd edition.
	
	\bibitem[Comunian et~al., 2014]{comunian2014training}
	Comunian, A., Jha, S.~K., Giambastiani, B.~M., Mariethoz, G., and Kelly, B.~F.
	(2014).
	\newblock Training images from process-imitating methods.
	\newblock {\em Mathematical Geosciences}, 46(2):241--260.
	
	\bibitem[Comunian et~al., 2012]{comunian20123d}
	Comunian, A., Renard, P., and Straubhaar, J. (2012).
	\newblock 3D multiple-point statistics simulation using 2D training images.
	\newblock {\em Computers \& Geosciences}, 40:49--65.
	
	\bibitem[Cressie, 1993]{Cressie:1993}
	Cressie, N. (1993).
	\newblock {\em Statistics for Spatial Data}.
	\newblock Wiley, New York, revised edition.
	
	\bibitem[Fabbri et~al., 2016]{Fabbri-et-al2016}
	Fabbri, P., Piccinini, L., Marcolongo, E., Pola, M., Conchetto, E., and
	Zangheri, P. (2016).
	\newblock Does a change of irrigation technique impact on groundwater
	resources? A case study in Northeastern Italy.
	\newblock {\em Environmental Science \& Policy}, 63:63--75.
	
	\bibitem[Fuglstad et~al., 2019]{fuglstad2019constructing}
	Fuglstad, G.-A., Simpson, D., Lindgren, F., and Rue, H. (2019).
	\newblock Constructing priors that penalize the complexity of Gaussian random
	fields.
	\newblock {\em Journal of the American Statistical Association},
	114(525):445--452.
	
	\bibitem[Galli et~al., 1994]{galli1994pros}
	Galli, A., Beucher, H., Le~Loc?h, G., Doligez, B., and Group, H. (1994).
	\newblock The pros and cons of the truncated Gaussian method.
	\newblock In {\em Geostatistical Simulations}, pages 217--233. Springer.
	
	\bibitem[Gelfand, 2000]{gelfand2000gibbs}
	Gelfand, A.~E. (2000).
	\newblock Gibbs sampling.
	\newblock {\em Journal of the American Statistical Association},
	95(452):1300--1304.
	
	\bibitem[Genz and Bretz, 2009]{GenzBretz2009}
	Genz, A. and Bretz, F. (2009).
	\newblock {\em Computation of Multivariate Normal and t Probabilities}.
	\newblock Lecture Notes in Statistics. Springer-Verlag, Heidelberg.
	
	\bibitem[Genz et~al., 2019]{GenzPackage}
	Genz, A., Bretz, F., Miwa, T., Mi, X., Leisch, F., Scheipl, F., and Hothorn, T.
	(2019).
	\newblock {\em {mvtnorm}: Multivariate Normal and t Distributions}.
	\newblock R package version 1.0-11.
	
	\bibitem[Le~Bl{\'e}vec et~al., 2017]{le2017modelling}
	Le~Bl{\'e}vec, T., Dubrule, O., John, C.~M., and Hampson, G.~J. (2017).
	\newblock Modelling asymmetrical facies successions using pluri-Gaussian
	simulations.
	\newblock In {\em Geostatistics Valencia 2016}, pages 59--75. Springer.
	
	\bibitem[Le~Bl{\'e}vec et~al., 2018]{le2018geostatistical}
	Le~Bl{\'e}vec, T., Dubrule, O., John, C.~M., and Hampson, G.~J. (2018).
	\newblock Geostatistical modelling of cyclic and rhythmic facies architectures.
	\newblock {\em Mathematical Geosciences}, 50(6):609--637.
	
	\bibitem[Liu et~al., 2019]{liu2019statistical}
	Liu, L., Shih, Y.-C.~T., Strawderman, R.~L., Zhang, D., Johnson, B.~A., Chai,
	H., et~al. (2019).
	\newblock Statistical analysis of zero-inflated nonnegative continuous data: A
	review.
	\newblock {\em Statistical Science}, 34(2):253--279.
	
	\bibitem[Marcotte and Allard, 2018]{marcotte2018gibbs}
	Marcotte, D. and Allard, D. (2018).
	\newblock Gibbs sampling on large lattice with GMRF.
	\newblock {\em Computers \& Geosciences}, 111:190--199.
	
	\bibitem[Mariethoz and Caers, 2014]{mariethoz2014multiple}
	Mariethoz, G. and Caers, J. (2014).
	\newblock {\em Multiple-point Geostatistics: Stochastic Modeling with
		Training Images}.
	\newblock John Wiley \& Sons.
	
	\bibitem[Martinetti and Geniaux, 2017]{martinetti2017approximate}
	Martinetti, D. and Geniaux, G. (2017).
	\newblock Approximate likelihood estimation of spatial probit models.
	\newblock {\em Regional Science and Urban Economics}, 64:30--45.
	
	\bibitem[Matheron et~al., 1987]{matheron1987conditional}
	Matheron, G., Beucher, H., De~Fouquet, C., Galli, A., Guerillot, D., Ravenne,
	C., et~al. (1987).
	\newblock Conditional simulation of the geometry of fluvio-deltaic reservoirs.
	\newblock In {\em {SPE} {A}nnual {T}echnical {C}onference and {E}xhibition}.
	Society of Petroleum Engineers.
	
	\bibitem[Pyrcz et~al., 2015]{pyrcz2015stratigraphic}
	Pyrcz, M.~J., Sech, R.~P., Covault, J.~A., Willis, B.~J., Sylvester, Z., Sun,
	T., and Garner, D. (2015).
	\newblock Stratigraphic rule-based reservoir modeling.
	\newblock {\em Bulletin of Canadian Petroleum Geology}, 63(4):287--303.
	
	\bibitem[Sartore et~al., 2016]{sartore2016spmc}
	Sartore, L., Fabbri, P., and Gaetan, C. (2016).
	\newblock {spMC}: an R-package for 3D lithological reconstructions based on
	spatial Markov chains.
	\newblock {\em Computers \& Geosciences}, 94:40--47.
	
	\bibitem[Simpson et~al., 2017]{simpson2017penalising}
	Simpson, D., Rue, H., Riebler, A., Martins, T.~G., S{\o}rbye, S.~H., et~al.
	(2017).
	\newblock Penalising model component complexity: A principled, practical
	approach to constructing priors.
	\newblock {\em Statistical Science}, 32(1):1--28.
	
	\bibitem[Strebelle, 2002]{strebelle2002conditional}
	Strebelle, S. (2002).
	\newblock Conditional simulation of complex geological structures using
	multiple-point statistics.
	\newblock {\em Mathematical Geology}, 34(1):1--21.
	
	\bibitem[Syversveen and Omre, 1997]{syversveen1997conditioning}
	Syversveen, A.~R. and Omre, H. (1997).
	\newblock Conditioning of marked point processes within a Bayesian framework.
	\newblock {\em Scandinavian Journal of Statistics}, 24(3):341--352.
	
	\bibitem[Tanner, 1996]{Tanner1996}
	Tanner, M.~A. (1996).
	\newblock {\em Tools for statistical inference: Methods for the Exploration of
		Posterior Distributions and Likelihood Functions}.
	\newblock Springer.
	
	\bibitem[Zhang, 2004]{zhang2004inconsistent}
	Zhang, H. (2004).
	\newblock Inconsistent estimation and asymptotically equal interpolations in
	model-based geostatistics.
	\newblock {\em Journal of the American Statistical Association},
	99(465):250--261.
	
\end{thebibliography}

\end{document}